# An integrated super-resolution THz 3D imaging system based on a linear nonlocal achromatic freeform Bessel beam lens and high-power oscillator–radiator array

Jin Chen[1#], Liang Gao[2#*], Hao Guo[1#*], Zhi Chao Chen[1#], Kang Jie Lin[2], Kam Man Shum[1], Ka Fai Chan[1], Chi Hou Chan[1,3*]

[1] State Key Laboratory of THz and Millimeter Waves, City University of Hong Kong, Hong Kong, 999077, China

[2] School of Information Science and Engineering, State Key Laboratory of Millimeter Waves, Southeast University, Nanjing, 210096, China

[3] Department of Electrical Engineering, City University of Hong Kong, Hong Kong, 999077, China

These authors contributed equally: Jin Chen, Liang Gao, Hao Guo, Zhi Chao Chen

[*]Corresponding author: haoguo8@cityu.edu.hk, lianggao@seu.edu.cn, eechic@cityu.edu.hk

## Abstract

High-performance terahertz (THz) 3D imaging is critical for non-destructive evaluation. However, conventional architectures are fundamentally limited by severe chromatic aberrations, modest spatial resolution, restricted depths of focus (DOF), and the bulky nature of commercial transceivers. While metasurfaces offer a compact alternative, achieving broadband achromatic super-resolution with an extended DOF remains a formidable challenge. Here, we present a highly integrated 3D THz imaging platform that synergizes a 3D-printed nonlocal freeform Bessel-beam lens with a high-power, 65-nm CMOS oscillator-radiator array. Harnessing nonlocal interactions within the lens, we generate an achromatic super-resolution Bessel beam (0.3–1 THz) with a sub-diffraction full width at half maximum (FWHM) of 0.65λ and a robust 4.7-mm DOF. Crucially, the system overcomes conventional sidelobe limitations, enabling high-fidelity 2D imaging of intricate sub-millimeter targets (e.g., USAF 1951 charts and QR codes) alongside robust 3D volumetric imaging through highly scattering media, such as printed circuit boards. By converging standard CMOS technology with additive manufacturing, this work establishes a versatile, cost-effective paradigm for next-generation integrated THz photonics.

## Introduction

Driven by the growing demands for high-speed communication[1-2], high-resolution imaging[3-4] of opaque objects, and biosensing of large molecules[5-6], there is a critical need to accelerate the development of high-performance terahertz (THz) imaging and sensing systems. While significant progress has been made in specific applications—such as cancer cell screening/quantification[7], phenotyping[8], and fruit ripeness sensing[9] -the realization of an integrated, high-performance THz system, particularly for imaging, remains elusive. Two primary bottlenecks impede the advancement of compact THz imaging architectures. First, unlike optical lenses, which have benefited from centuries of refinement, THz counterparts remain comparatively underdeveloped, with high-performance designs appearing only sporadically in literature. Second, most current systems rely on bulky, expensive benchtop instruments, such as THz Time-Domain Spectroscopy (TDS) and frequency extender modules. These non-integrable components severely hinder the widespread deployment and commercial viability of THz technology.

To overcome these inherent limitations, the development of high-performance THz lenses—pivotal components that dictate overall imaging quality—has become a research priority. Given that THz optics has historically evolved in parallel with classical optics, where dielectric refractive lenses predominate, THz refractive lenses were among the first to be studied and widely deployed. However, the significantly longer wavelengths of THz radiation compared to visible light necessitate refractive elements that are prohibitively bulky. Furthermore, conventional refractive lenses are frequently plagued by severe chromatic aberration, subpar resolution, and a constrained DOF. Reconciling the demand for miniaturization with the need for enhanced imaging performance constitutes a major unresolved issue. The advent of metasurfaces has introduced a transformative paradigm for designing ultrathin, high-performance optics. While periodic nanopillar-based metasurfaces have enabled a diverse range of optical functionalities, including achromatic[10-12] and super-resolution[13] metalenses, holographic displays[14-16], and structured light generators[17-19], their THz counterparts[20-23] have primarily focused on replicating optical phenomena. Consequently, the distinct technical requirements of practical THz imaging systems have not been adequately addressed in prior research.

Our previous research on aberration-free THz metalenses[3] for wide-field super-resolution imaging addressed critical practical requirements by employing gradient index (GRIN) architectures. The exceptional imaging performance achieved underscores the significant potential of GRIN designs for high-performance THz metalenses. Unlike conventional optics, the primary advantage of THz radiation lies in its ability to penetrate dielectric objects to a certain depth, necessitating the development of achromatic, super-resolution THz metalenses with an extended DOF for 3D imaging. However, simultaneously achieving achromaticity, super-resolution, and a large DOF within a GRIN-based framework still constitutes a major hurdle.

Owing to its inherent self-healing and non-diffractive properties, the THz Bessel beam is uniquely suited for achromatic super-resolution 3D imaging. Given that resolution is intrinsically linked to the cone angle, super-resolution performance demands a large cone angle. While nanopillar-based metasurfaces are standard for phase modulation, their resonant behavior impairs achromatic performance and shortens DOF at large cone angles. Consequently, transforming the phase profile into a gradient-index (GRIN) medium of specific thickness offers an alternative; however, the stringent fabrication requirements associated with high-index-gradient metamaterials of a large cone angle severely impede their practical implementation.

For the miniaturization of THz sources, transistor circuits based on standard semiconductor processes, such as 65-nm CMOS, are prioritized. However, the maximum oscillation frequency of silicon-based transistors in standard 65-nm CMOS processes is typically restricted to approximately 0.3 THz using conventional design methodologies; the output power of THz radiators employing a high-order harmonic oscillator at higher frequencies is severely constrained. THz radiation exceeding 0.6 THz—doubling the intrinsic $f_{max}$ of the 40-nm CMOS—was realized through integration of self-feeding third-harmonic oscillators and a dual-mode (in-phase coplanar waveguide (CPW) and out-of-phase slot line) coupling technique[25]. Nevertheless, such a design is only one-dimensional, and the resulting total radiated power remains modest at −16.1 dBm. Consequently, achieving sufficient output power beyond 0.6 THz represents a major technical bottleneck for standard 65-nm CMOS-based integrated systems.

This paper proposes tailored strategies to establish an integrated super-resolution 3D THz imaging platform that effectively overcomes the performance bottlenecks inherent in current THz lenses and CMOS-based radiators. To reconcile the demand for high resolution with a relatively low refractive index, a nonlocal mechanism is synergistically incorporated into the GRIN profile. Given the technical specifications of high-precision 3D printing for large-aperture gradient structure and the previously proven equivalence between metalenses and freeform lenses[4], this nonlocal GRIN profile is translated into a corresponding freeform lens for experimental realization. An achromatic super-resolution Bessel beam across the 0.3–1 THz band (FWHM = 0.65λ and DOF = 4.7 mm) could be generated. Building on our prior research in 2D scalable coupled-oscillator arrays[24-27], we report a 288-element 0.67 THz source array in 65-nm CMOS delivering 12.94 dBm radiated power and 37.47 dBm lensless effective isotropic radiated power (EIRP). The resultant integrated 3D imaging system, incorporating the custom linear nonlocal freeform lens, operates at 0.658 THz. The corresponding imaging results demonstrate that sub-millimeter features in intricate patterns, such as the USAF 1951 resolution chart and QR codes, could be resolved with a precision of 0.2 mm. Ultimately, this breakthrough transcends the conventional boundaries of Bessel beams, unlocking their capability for high-resolution imaging of complex structures. Furthermore, 3D imaging at different depths could also be detected. These results validate the feasibility of constructing cost-effective, high-performance THz imaging platforms through the synergistic integration of standard CMOS processes and 3D printing technology.

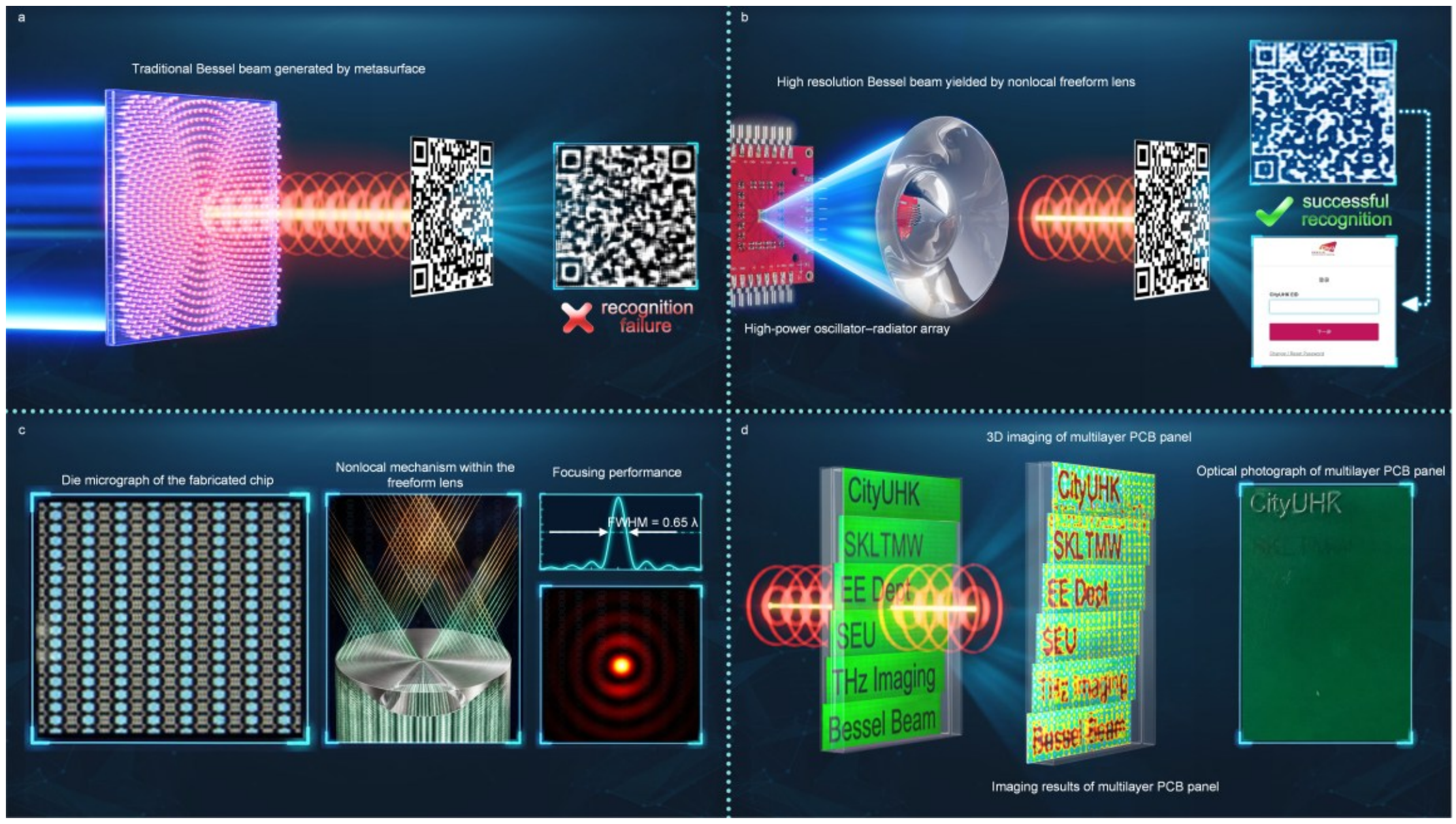

**Fig. 1. The overall performance of an integrated super-resolution THz imaging system comprising a linear nonlocal Bessel-beam lens and a high-power radiator array.** a. Traditional Bessel beams fail to discern the high-spatial-frequency complex pattern, such as a QR code. b. THz signals generated by the oscillator–radiator array illuminate the designed nonlocal freeform lens, yielding the high-resolution Bessel beam to decode the QR code successfully. c. The details of the designed oscillator–radiator array, the nonlocal mechanism of our proposed freeform lens, and the corresponding focusing performance. d. 3D imaging performance for multi-layered PCB.

# Results

## Theoretical design of THz linear nonlocal freeform Bessel beam lens

The phase distribution depicted in Fig. 2a to yield a Bessel beam is explained as follows:

$$\varphi(r) = kr\sin\theta \tag{1}$$

In which $k$ is the wavevector in the air and θ represents the cone angle of the expected Bessel beam. Furthermore, Bessel beams constitute a family of solutions to the Helmholtz equation, and the FWHM of the transverse field distribution for the zero-order Bessel beam is derived as:

$$\mathrm{FWHM} = \frac{0.358\lambda}{\sin\theta} \tag{2}$$

Where $\lambda$ denotes the wavelength in the air, the details of the corresponding mathematical derivation are provided in Note 1 of the supplementary file. Equation 2 indicates that the resolution of the Bessel beam is proportional to its cone angle.

Metasurfaces engineered to implement the phase profile required for Bessel beam synthesis, as defined in Eq. (1), are widely recognized as meta-axicons [28–31]. However, the intrinsic resonant nature of nanopillar unit cells typically constrains the bandwidth over which high-fidelity Bessel beams can be sustained—a limitation that becomes increasingly pronounced at steep cone angles. Consequently, reports on their practical deployment in imaging applications remain scant.

Enlightened by our prior experience in realizing aberration-free metalens[3] by GRIN profile, it is natural to transform the phase profile of the Bessel beam into the corresponding GRIN distribution illustrated in Fig. 2a:

$$n(r) = n_0 - \frac{r}{d}\sin\theta \tag{3}$$

In which $n_0$ defines the largest value of the refractive index at the center of the lens, d symbolizes the thickness of the GRIN lens. From a practical implementation standpoint, the refractive index profile $n(r)$ of a GRIN lens must remain greater than unity. As established in Eq. (2), the spatial resolution of the synthesized Bessel beam is governed by the cone angle θ, which in turn necessitates a concomitant increase in the base index $n_0$ and the radial profile $n$(r). Consequently, achieving the desired super-resolution performance dictates a high-refractive-index GRIN distribution. However, materials that possess both a high refractive index and the processability required for complex gradient metamaterials are exceptionally scarce, posing significant fabrication challenges, particularly in the THz regime. This fundamental trade-off between super-resolution Bessel beams and the necessity for low-refractive-index GRIN profiles within conventional design paradigms has historically stymied the broad adoption of Bessel beams in practical super-resolution imaging applications.

To address the above fundamental dilemma, a nonlocal mechanism is introduced to the theoretical framework of the GRIN lens. In consequence, the nonlocal GRIN distribution sketched in Fig. 2a for the Bessel beam is written as follows:

$$n(r) = n_0 - \frac{|r - 0.5R|}{d}\sin\theta \tag{4}$$

Where $R$ is the radius of the GRIN lens, to make an effective comparison between a nonlocal and a local GRIN lens for Bessel beam generation, the values of $n_0$, the radius $R$, and the thickness $d$ in equations 3 and 4 are unified as 1.6, 7.5mm, and 3.75mm, respectively. Subsequently, the cone angles of the local and nonlocal lenses are deduced to be arcsin(0.3) and arcsin(0.6), respectively, implying that the resolution of the Bessel beam generated by the nonlocal lens is much larger than that of the local one. To further demonstrate that the nonlocal mechanism will significantly enhance the Bessel beam's resolution, ray tracing and electric field distributions of the nonlocal lens and the local lens are calculated as shown in Fig. 2b, Fig. S1, and Fig. S2, respectively.

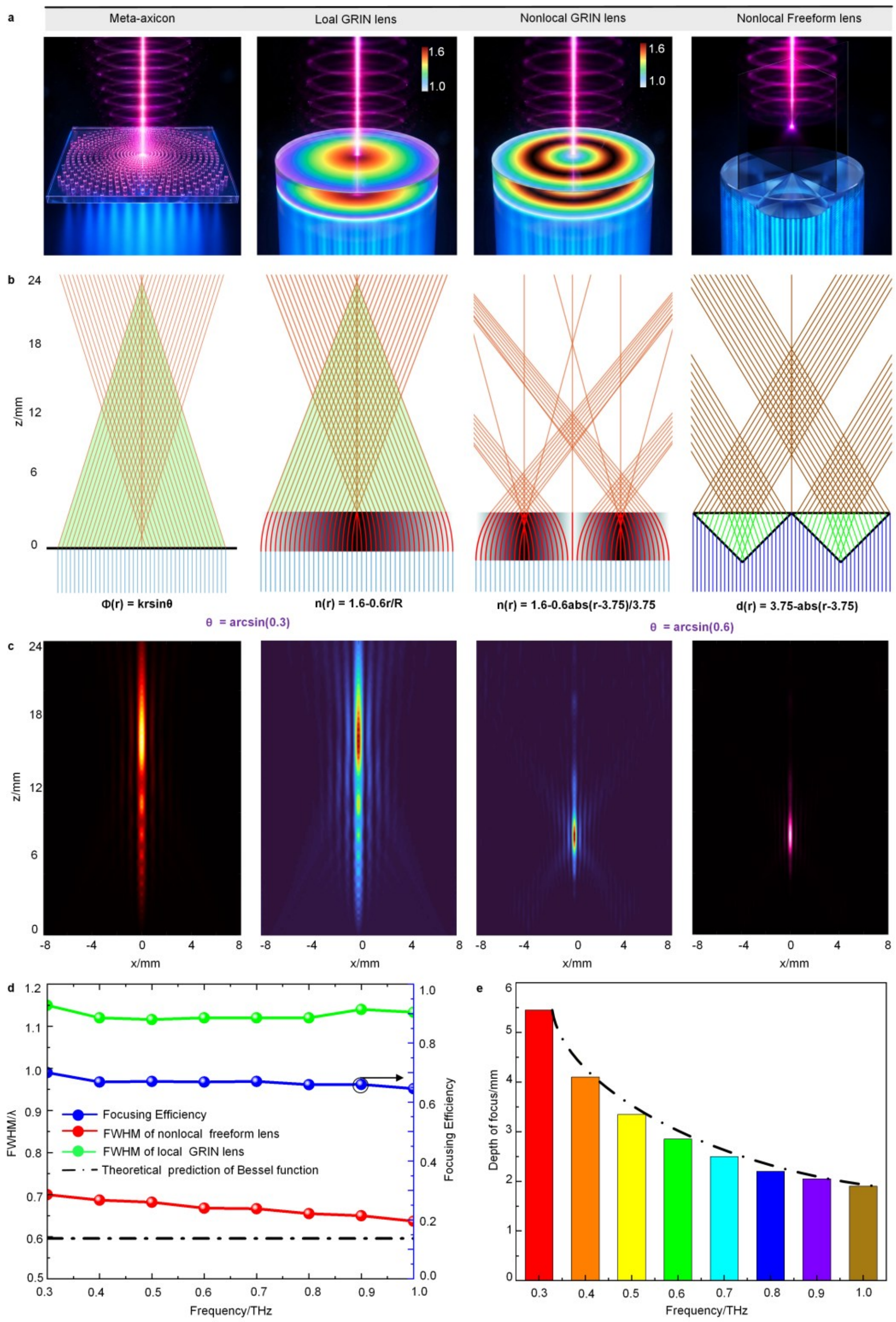

**Fig. 2. Theoretical design of super-resolution nonlocal freeform Bessel beam THz lens. a.** Evolution of the Bessel beam lens from meta-axion to nonlocal freeform lens. **b-c** Calculated ray tracing and electric field distributions at 0.7 THz of four kinds of Bessel beam lenses. **d**. Derived focusing efficiency and FWHM from Fig. 2b. **e**. Depth of focus over the 0.3–1 THz band.

A comparative analysis of the calculated FWHM for local and nonlocal GRIN lenses is presented in Supplementary Fig. S3. Both ray-tracing simulations and electric field distributions reveal that the nonlocal mechanism enhances the spatial resolution of the Bessel beam by widening the effective cone angle. Notably, the Bessel beam generated by the proposed nonlocal GRIN lens exhibits robust achromatic performance across the 0.3–1.0 THz band. However, realizing gradient metamaterials that precisely match the target GRIN profile over a 15-mm aperture at 0.6 THz remains technically prohibitive with state-of-the-art high-precision 3D printing. To circumvent this, we leverage the equivalence between refractive-index (GRIN) and thickness (freeform) profiles established in our prior study[4]. Consequently, the nonlocal GRIN lens is functionally replaced by its equivalent freeform counterpart, as illustrated in Fig. 2a, and this nonlocal freeform lens is manufactured by 3D printing.

To evaluate the performance of the proposed nonlocal freeform lens, the calculated electric field distributions in the x-z and x-y planes from 0.3 to 1.0 THz are presented in Figs. S4 and S5, respectively. These results confirm the successful generation of achromatic Bessel beams and validate the efficacy of the GRIN-to-freeform equivalence in the nonlocal regime. The corresponding FWHM and focusing efficiencies are summarized in Fig. 2c. While the theoretical resolution (dashed black line) represents an asymptotic limit derived from the ideal Bessel function, the FWHM achieved by our nonlocal freeform lens (dashed red line) demonstrates a near-ideal focusing response. Due to the inherent energy distribution into the Bessel beam's sidelobes, the focusing efficiency remains approximately 60% across the entire operational bandwidth. To further characterize the super-resolution performance, the DOF and the corresponding x-y plane electric field distributions at 0.7 THz are illustrated in Figs. 2d and 2e. As the operating frequency scales from 0.3 to 1.0 THz, the DOF undergoes a reduction from 5.4 mm to 2.0 mm. Nevertheless, the spatial resolution remains invariant throughout the axial range, validating the efficacy of the nonlocal freeform lens for high-fidelity 3D imaging.

**Experimental demonstration of the proposed THz nonlocal freeform lens**

The experimental configuration used to map the spatial electric-field distribution and characterize the achromatic Bessel beam was adapted from our previously established setup[4], with a comprehensive schematic provided in Fig. S6. As illustrated by the measured electric-field intensity profiles in the *x-z* and *x-y* planes (Fig. 3a), a broadband achromatic Bessel beam is experimentally realized across the 0.3 to 1.0 THz range. Furthermore, the measured FWHM and focusing efficiency agree well with theoretical predictions across the entire operating bandwidth (Fig. 3b), with minor discrepancies attributed primarily to experimental uncertainties. Notably, owing to the inherent material dispersion of the constituent resin (Fig. S7), the focal length is not strictly invariant; rather, it exhibits a positive correlation with the incident frequency, as further captured in Fig. 3c. Furthermore, the dispersion of the resin also contributes to the obvious differences between the theoretical DOF and the measured one, as depicted in Fig. 3 d. The measured DOF remained around 4.0- 5.0 mm, indicating that the fabricated nonlocal freeform lens is very useful for 3D imaging. To further demonstrate the capability of 3D imaging, the measured FWHM and the normalized intensity profile along the propagation distance at 0.7 THz are extracted and displayed in Fig. 3e, revealing that the FWHM remains stable around 0.65-0.75$\lambda$ in the whole DOF. In other words, the corresponding imaging resolution also remains stable within the DOF.

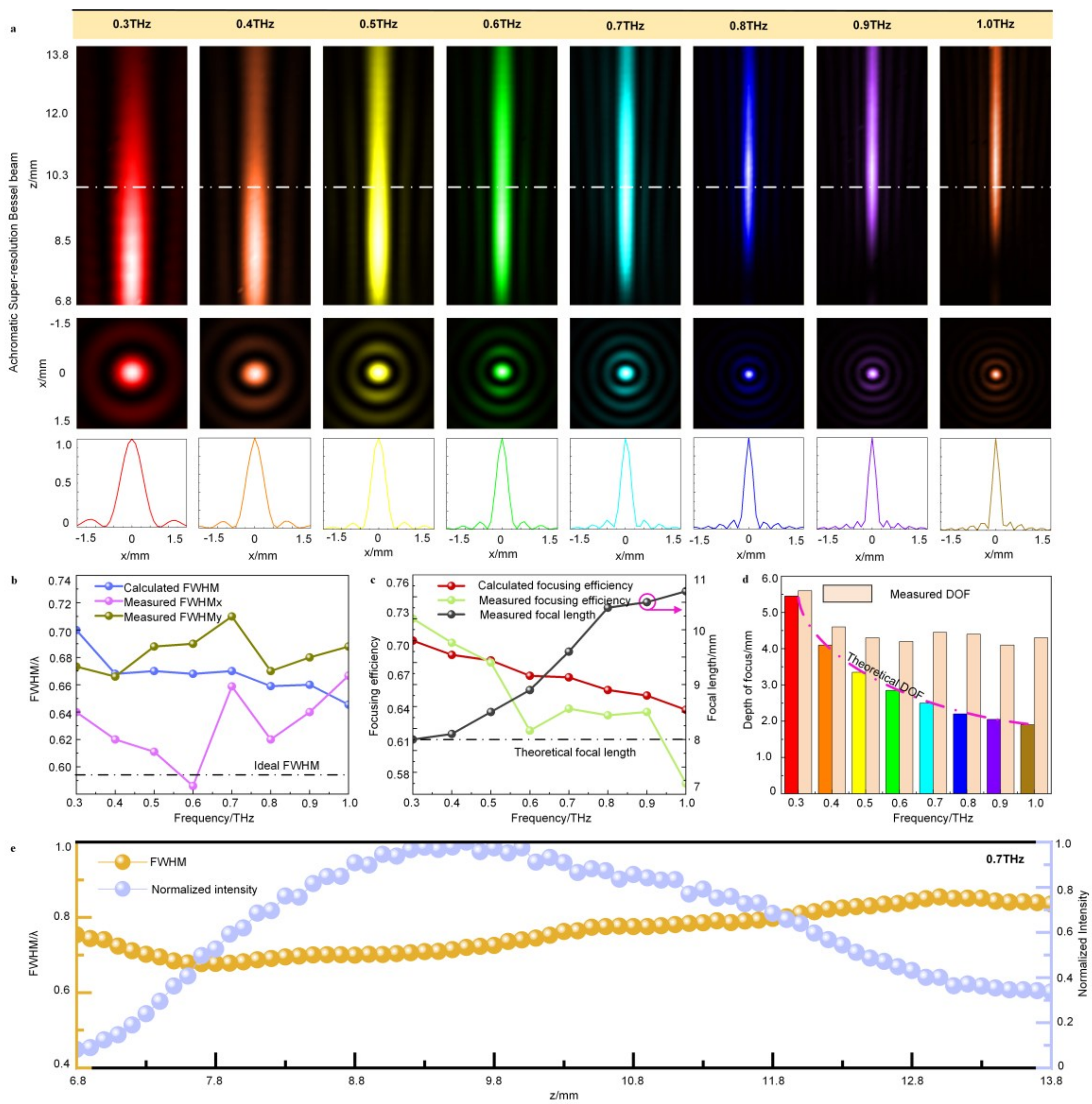

**Fig. 3. Experimental achromatic Bessel beam performance of THz nonlocal freeform lens. a.** Measured results of electric field distributions in the x-z plane and x-y plane from 0.3 to 1.0 THz. **b-c**. Comparison between the measured FWHM, focusing efficiency, focal length, and those of the theoretical one. **d**. Comparison between the measured DOF and the theoretical one. **e.** Measured FWHM and intensity of the THz needle's main lobe along the propagation direction at 0.7 THz.

## The design of the THz oscillator–radiator array at 0.67THz based on 65nm CMOS

Harmonic oscillators are widely employed for THz signal generation. By pushing the fundamental oscillation frequency above 0.2 THz, the third-harmonic output can be extended beyond 0.6 THz. However, such high operating frequencies typically encounter degraded harmonic conversion efficiency, which severely limits the output power of a single oscillator and hinders its utility in power-demanding applications such as THz imaging. Consequently, multi-element coherently coupled array architectures have emerged as a vital paradigm for scaling THz output power. Fig. 4a illustrates the proposed coherent THz source array. This architecture leverages scalable coupled oscillators to achieve fundamental-frequency synchronization across the array, while the third-harmonic signals from individual elements are extracted and radiated by on-chip antennas. Operating in a pre-designed coupling mode ensures that the harmonic signals from all elements are strictly in-phase. This facilitates spatially coherent power combining, yielding a highly directive beam with a substantial EIRP.

The core of this coherent THz radiator array lies in its scalable coupling architecture and its holistic co-design with the on-chip antennas. As depicted in Fig. 4b, the proposed coupling topology enforces an out-of-phase condition between adjacent elements along the rows and an in-phase condition along the columns, rigorously maintaining the requisite phase relationships. Specifically, the row-wise out-of-phase coupling facilitates localized power combining between adjacent elements, whereas the column-wise in-phase coupling guarantees global coherent combining across the entire array. Furthermore, the collaborative design of the coupling network and antenna structure significantly minimizes the inter-element spacing. This compact footprint not only suppresses grating lobes—thereby enhancing the array's radiation profile—but also maximizes the number of integrated elements within a given die area, ultimately boosting the total radiated power.

As illustrated in Fig. 4c, the $2 \times 2$ multifunctional radiating unit cells are built upon the architecture reported in [32]. The oscillator, coupling network, and antenna are co-optimized to deliver high output power within a strictly compact layout. The gate and source networks are engineered to provide optimal oscillator-embedding impedance while facilitating array coupling; specifically, the gate network supports the out-of-phase mode, whereas the source network maintains the in-phase mode. Concurrently, the drain network accommodates both the fundamental oscillation conditions and the DC bias. Crucially, it serves as an efficient third-harmonic extraction pathway, directly feeding the harmonic signals to the antennas for radiation.

By tiling this multifunctional unit, a two-dimensional, scalable array can be synthesized. However, scaling the array to hundreds of elements introduces profound design challenges. Large-scale oscillator arrays are susceptible to multi-mode oscillation, risking resonance in undesired coupling modes that compromise the target frequency and output power. Furthermore, for arrays exceeding a hundred elements, perfect unit-to-unit uniformity is fundamentally precluded by process, voltage, and temperature (PVT) variations. These non-idealities can trigger spurious modes and introduce inter-element phase errors, which severely degrade spatial power-combining efficiency. Consequently, the design of a large-scale array necessitates rigorous verification of mode stability and careful management of peripheral boundary conditions to ensure robust oscillator operation.

The proposed large-scale radiator array was fabricated in a standard 65-nm CMOS process. As depicted in the chip micrograph in Fig. 4d, the total die area is 2.55 mm×2.35 mm. By systematically sweeping the supply and bias voltages, the frequency tuning range of the array was characterized, yielding the results plotted in Fig. 4e. The corresponding DC power consumption across the tuning range is detailed in Fig. 4f. To accurately evaluate the array's spatial power-combining performance, the free-space and instrumental link losses from the chip to the signal analyzer were rigorously calibrated following the methodology outlined in [25]. Accounting for these calibrated losses, the measured EIRP over the operating band is shown in Fig. 4h, with a peak of 37.47 dBm at 0.6707 THz. At this peak frequency, the E-plane and H-plane radiation patterns were acquired by mounting the device under test (DUT) on a motorized rotation stage with a 1° angular resolution, as shown in Fig. 4g. The antenna directivity, derived from the measured orthogonal 2D patterns using the numerical integration method described in [34], evaluates to 22.05 dBi. This value shows a noticeable discrepancy with the simulated 3D directivity of 24.53 dBi. Because total radiated power is calculated by subtracting the directivity from the EIRP (in decibels), adopting a higher directivity yields a lower calculated power. Therefore, to provide a stringent and conservative assessment of the array's performance, the higher simulated directivity is employed to extract the radiated power from the measured EIRP, as plotted in Fig. 4i. Consequently, the array achieves a peak radiated power of 12.94 dBm at 0.6707 THz, which translates to a peak DC-to-THz radiation efficiency of 0.374%.

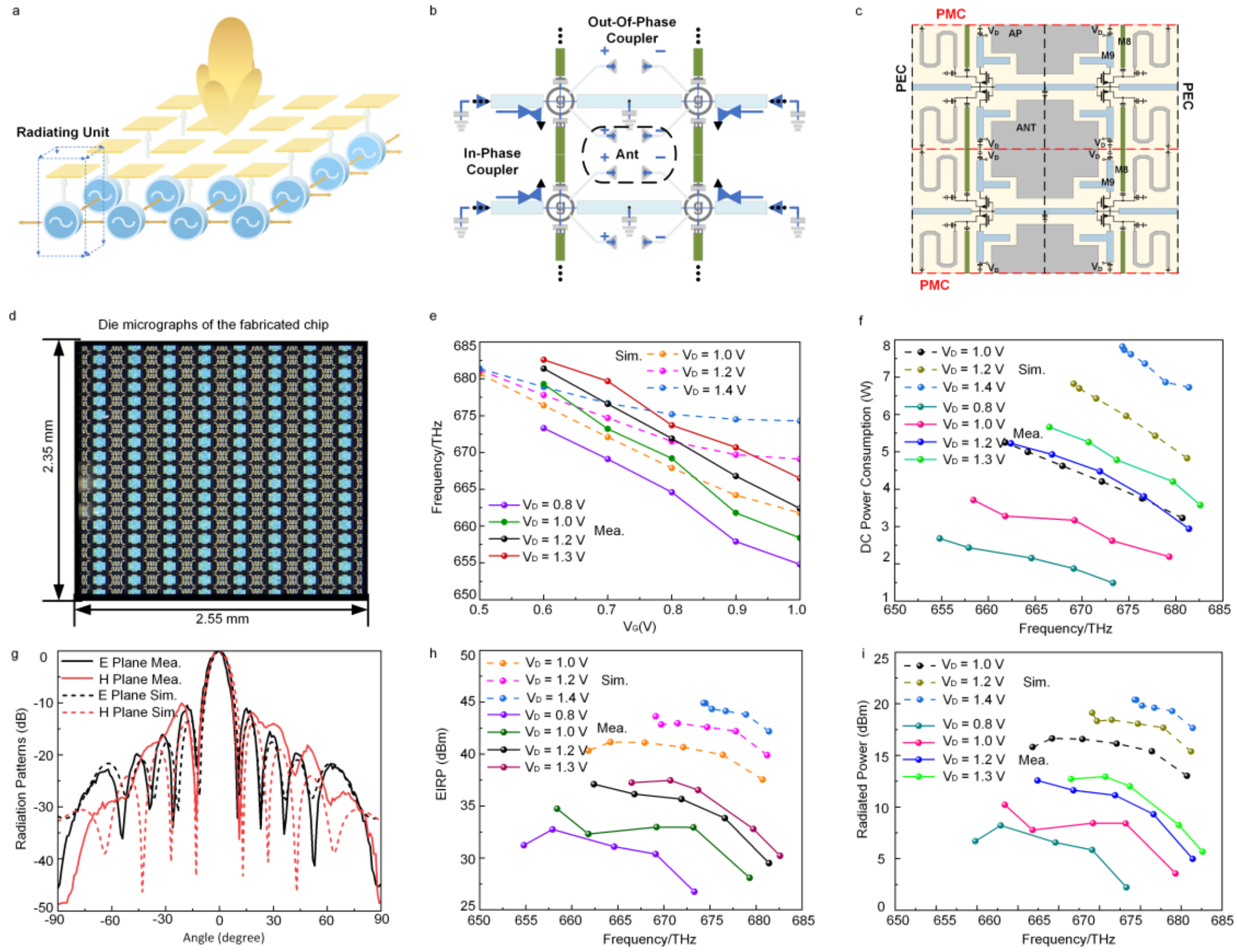

**Fig. 4 Design and demonstration of THz oscillator–radiator array. a.** Architecture of the coherent THz source array. **b.** Scalable coupled oscillator array topology. **c**. Schematic of the 2 × 2 radiating unit. **d**. The die micrographs of the fabricated chip. **e**. Simulated and measured frequency tuning range. **f**. Simulated and measured DC power consumption. **g**. Simulated and measured radiation pattern at 0.6707 THz. **h**. Simulated and measured EIRP. **i**. Simulated and measured radiated power.

## The integrated 3D THz imaging system of a nonlocal freeform lens and an oscillator–radiator array

Based on the proposed nonlocal freeform lens and oscillator-radiator array, we construct the compact imaging system depicted in Fig. 5a. The oscillator-radiator array emits *y*-polarized THz waves from 0.65 to 0.67 THz. Operating the array at maximum capacity introduces significant challenges regarding thermal dissipation and long-term output stability. More critically, prolonged full-power operation leads to a significant degradation in overall system performance. Consequently, to mitigate excessive power consumption during prolonged operation, only one-third of the available channels were activated for imaging at 0.658 THz. The corresponding output power is given in Fig. S11. These incident waves are transformed by the nonlocal freeform lens into a super-resolution Bessel beam with an extended DOF onto the sample. The THz signals passing through the sample are subsequently collected by a receiving probe. The system size can be reduced further using the 0.68 THz receiver IC using 65-nm CMOS[35]. To acquire 2D imaging data, the sample is mounted on a motorized stage for raster scanning in the *x-y* plane. Finally, the measured 2D intensity data are post-processed to yield the imaging results. The intrinsic sidelobes of conventional Bessel beams often introduce imaging artifacts, severely degrading the resolution of complex patterns. To validate the superior imaging fidelity of the proposed nonlocal freeform lens, a USAF 1951 resolution test chart and QR codes with feature sizes of 0.2 mm and 0.3 mm were employed as imaging targets, as illustrated in Fig. 5b. The USAF 1951 resolution chart is produced by selectively removing material from a thin aluminum sheet, and the QR codes are embedded within the inner layer of the thin PCB panel via lithographic patterning. The actual image of the corresponding PCB panel is shown in Fig. S12. Although pronounced artifacts are evident in the raw imaging results (Fig. 5c), fine details of both the USAF 1951 resolution target and the embedded QR code are distinctly resolved. Notably, the QR code featuring a minimum critical dimension of 0.3 mm remains machine-readable; even without image post-processing, it can be successfully scanned by a smartphone, directing users to the login homepage of City University of Hong Kong, indicating the robust imaging capability of the proposed nonlocal Bessel beam lens for complex patterns. To further enhance imaging quality, deconvolution, stripe filtering, and axial constraint are applied to the raw data. The corresponding details of these post-processing technologies are provided in Supplementary Note 3. As depicted in Fig. 5d, the applied post-processing substantially mitigates the artifacts present in the raw imaging data, an improvement particularly evident in the USAF 1951 resolution chart. Notably, while the raw THz image of the QR code with a 0.2-mm feature size (which encodes the URL of the Web of Science, used solely as an imaging target) remains unreadable to standard smartphone scanners, the post-processed image is successfully decoded, reliably directing users to the URL of the Web of Science. Similarly, for the 0.3-mm QR code, the post-processing procedure yields a significantly enhanced image contrast compared to the raw data. Crucially, the minimum resolvable feature size for both the USAF chart and the QR codes converges at 0.2 mm, thereby establishing that the ultimate spatial resolution of our compact THz imaging system for complex patterns is 0.2 mm.

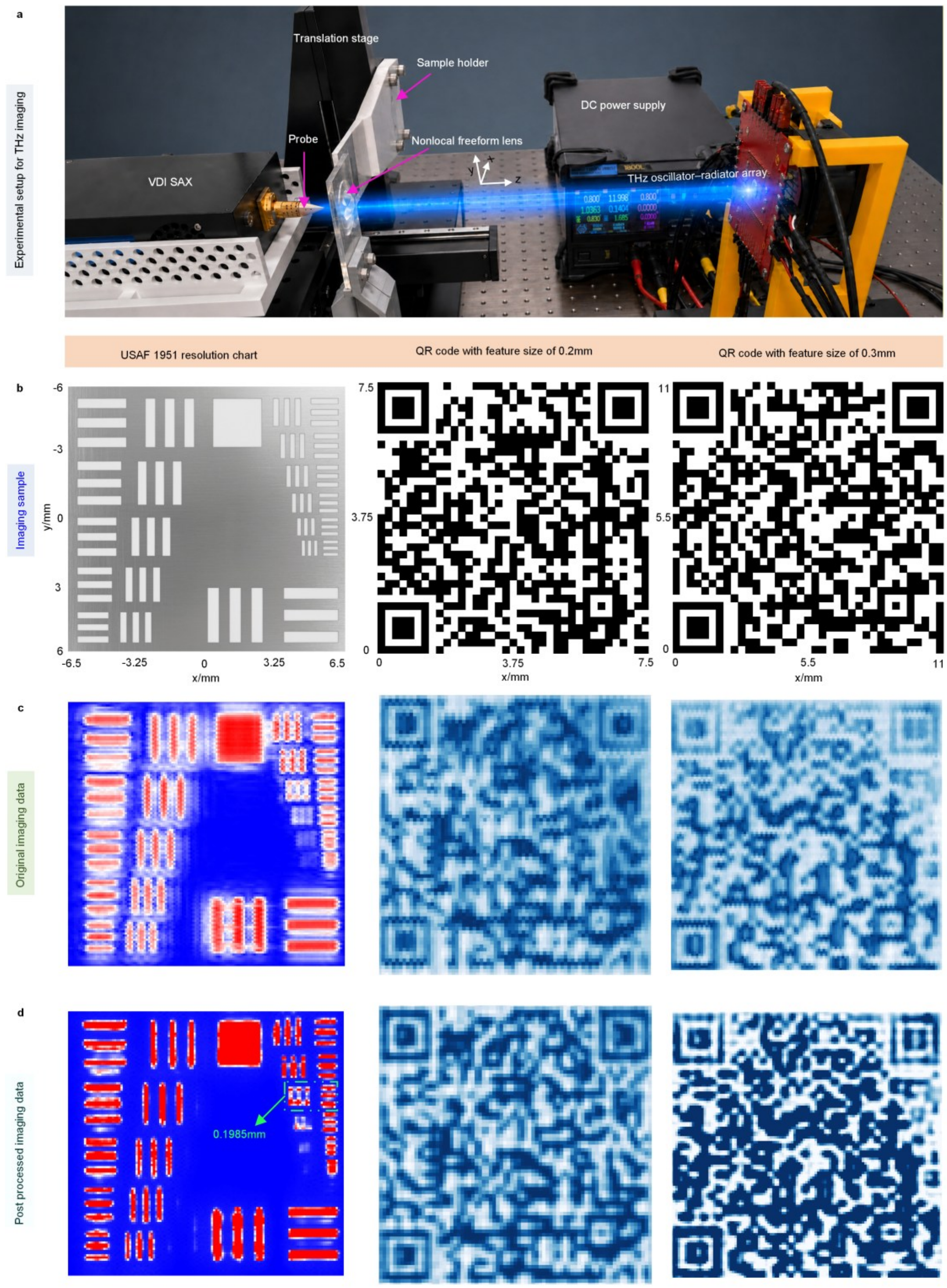

Fig. 5. **The integrated 3D THz imaging system and the corresponding imaging results**. **a**. The integrated 3D THz imaging system established by a nonlocal freeform lens and an oscillator–radiator array. **b**. Imaging samples of the USAF 1951 resolution chart and QR codes with a feature size of 0.2 mm and 0.3 mm. **c**. Raw data of imaging results. **d**. Post-processed imaging results.

On the other hand, the self-healing property of the Bessel beam enables 3D imaging at different depths within the scattering medium, such as a PCB panel. To further validate the three-dimensional imaging capability of the developed THz system, specific text patterns with 0.2-mm-wide (i.e., CityUHK, SKLMTW, EE Dept, SEU,

THz Imaging, and Bessel Beam) were embedded at distinct axial depths within a 1.5-mm-thick PCB substrate. A two-dimensional (2D) optical photograph of the multi-layered PCB substrate is presented in Fig. 6a, with its corresponding 3D geometric layout illustrated in the perspective view of Fig. 6b. In the raw volumetric imaging results depicted in Fig. 6c, the uppermost four layers of text are distinctly discernible. However, the deepest two layers (i.e., 'THz Imaging' and 'Bessel Beam') suffer from significant blurring. This degradation is primarily attributed to strong scattering effects induced by the heterogeneous textile structure of the PCB substrate. To retrieve these obscured features, a customized post-processing pipeline—comprising deconvolution, stripe filtering, and axial constraints—is applied to the raw dataset. As demonstrated in Fig. 6d, this integrated post-processing substantially suppresses the scattering artifacts, enabling all six layers of text to be resolved with high fidelity.

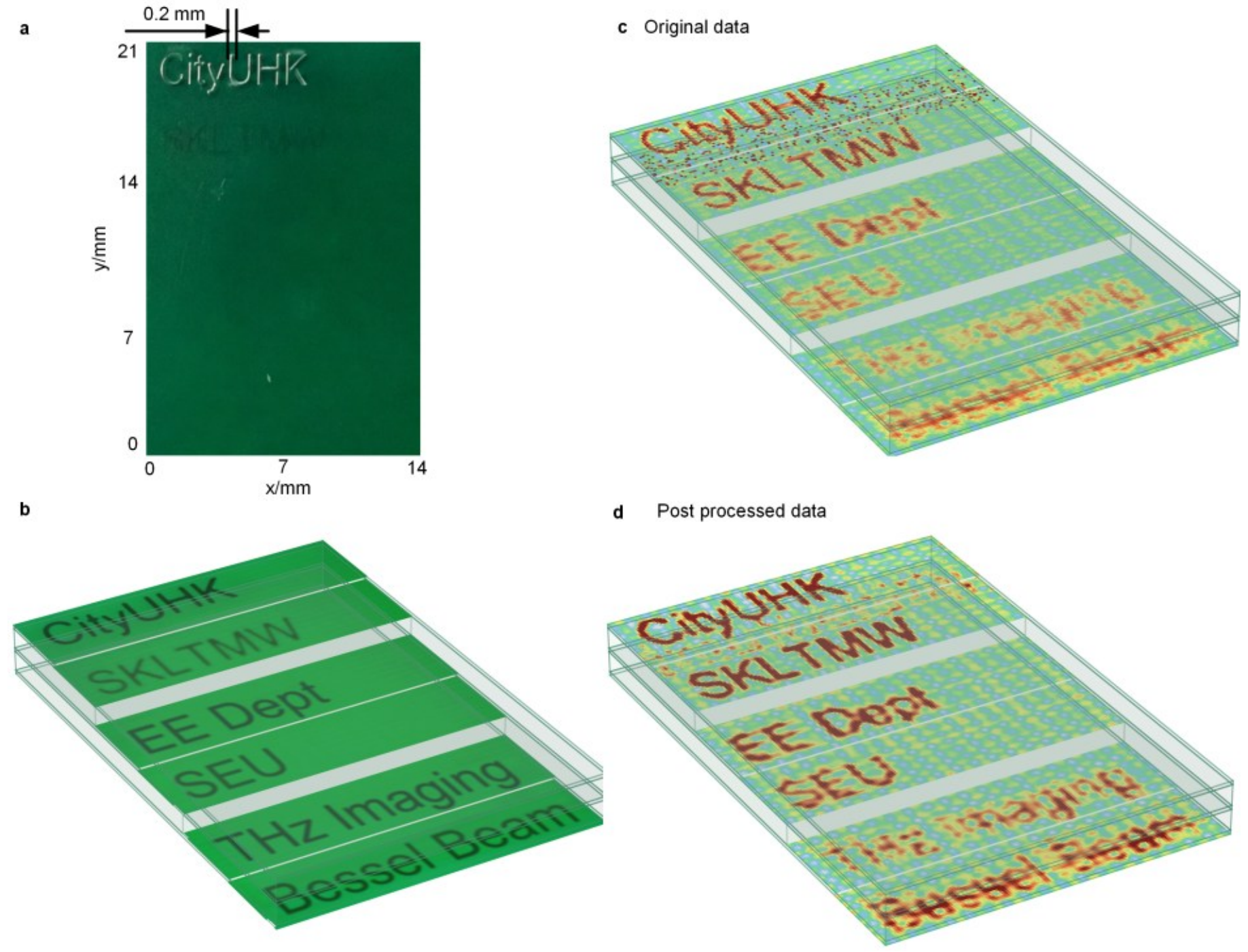

Fig. 6. **Multi-layer PCB and the corresponding imaging results**. **a**. Optical photograph of the multi-layered PCB substrate. **b**. 3D geometric layout of the multi-layered PCB substrate. **c**. Raw data of imaging results. **d**. Post-processed imaging results.

Finally, to experimentally validate the maximum penetration depth at which the Bessel beam fully reconstructs and retrieves imaging information beyond the PCB dielectric layer, a 2.0 mm-thick PCB substrate with specific text patterns with 0.5-mm-wide (i.e., CityUHK, SEU, SKLMTW, and Prof CHAN) embedded at distinct axial depths was selected as the imaging sample. A 2D optical photograph and the corresponding 3D geometric layout of the multi-layered PCB substrate are provided in Figs. S13a and S13b, respectively. Upon examining the raw tomographic data (Fig. S13c), the uppermost two text layers are distinct; however, the deeply embedded two layers (i.e., SKLMTW and Prof CHAN) are severely blurred by scattering. Following the application of a comprehensive post-processing pipeline (deconvolution, stripe filtering, and axial constraints), the scattering artifacts are substantially mitigated. Consequently, as shown in Fig. S13d, the top three layers are all successfully reconstructed and resolved with remarkably high fidelity. Consequently, the maximum penetrative depth at which the Bessel beam can successfully reconstruct and retrieve structural information through the PCB dielectric layer is firmly established at 2 mm. To benchmark performance under these extreme boundary conditions, we compared the imaging results obtained using our custom-designed oscillator-radiator array with those from a state-of-the-art commercial VDI extender module (Figs. S13c–f). Notably, the comparison clearly demonstrates that the THz signals generated by our integrated array exhibit superior imaging quality and signal integrity, markedly outperforming the commercial VDI counterpart.

## Discussion

In summary, we have demonstrated an integrated THz imaging platform that synergizes a 3D-printed nonlocal freeform Bessel-beam lens with a high-power, 65-nm CMOS-based oscillator-radiator array. By engineering nonlocal interactions within the linear unit cells of the lens, we achieved a super-resolution Bessel beam, delivering a sub-diffraction FWHM of 0.65λ and a valid DOF of 4.7 mm. Crucially, the system overcomes the inherent sidelobe limitations of conventional Bessel beams, enabling high-fidelity 2D imaging of intricate, sub-millimeter structures—such as USAF 1951 charts and QR codes—with a remarkable ultimate spatial resolution of 0.2 mm. Furthermore, by harnessing the intrinsic self-healing properties of the Bessel beam, we demonstrate deep-tissue 3D volumetric imaging within highly scattering media, successfully resolving 0.2-mm-wide text embedded inside a multi-layered PCB substrate at 0.658 THz. The consistent achievement of a 0.2-mm resolving limit across both surface and deep-penetration tomographic imaging strictly validates the exceptional robustness and resolving capability of our proposed THz platform. Ultimately, by bridging standard CMOS technology with additive manufacturing, our work establishes a highly versatile and cost-effective paradigm for next-generation, high-performance integrated THz systems.

## Methods

**Verification of mode stability and management of peripheral boundary conditions**

Figure S8 depicts the schematic of the proposed 18 × 16 array, which doubles the scale of the design reported in [32]. To mitigate the challenges, the gate terminals are independently biased in groups of three rows, yielding six distinct gate tuning voltages ($V_g$). Additionally, large decoupling capacitors are implemented at the gates of the peripheral units to establish a virtual ground. This creates an equivalent perfect electric conductor (PEC) boundary, ensuring that the edge elements share the same electromagnetic boundary conditions as the interior oscillators. Co-simulation of the full array with these boundary capacitors under mismatch conditions yields the waveforms shown in Fig. S9. The results confirm that the array robustly sustains the desired coupling mode, thereby validating the proposed topology. Furthermore, fine-tuning the six independent gate-bias voltages allows for the adjustment of the free-running frequency of each sub-array group. As demonstrated in[33], this capability enables the array to re-establish a globally synchronized, stable coupling state, providing a crucial degree of tuning freedom to counteract severe PVT variations.

**Experimental measurements of THz oscillator–radiator array**

Fig. S10 illustrates the experimental setup to measure the output frequency and the EIRP. To strictly satisfy far-field measurement conditions, a spatial separation of 1.15 m was maintained between DUT and the receiver. The radiated signals were captured by a VDI WR1.5 extension module (PXAX) equipped with a standard horn antenna and subsequently analyzed using an Agilent N9030A signal analyzer.

**Data availability**

The data that support the findings of this study are available from the corresponding authors upon reasonable request.

**Code availability**

The data that support the findings of this study are available from the corresponding authors upon reasonable request.

## Acknowledgments

This research was supported in part by the Research Grants Council of the Hong Kong under the General Research Fund (grant CityU 11214123).

Author information

These authors contributed equally: Jin Chen, Liang Gao, Hao Guo, Zhi Chao Chen

Authors and affiliations.

**State Key Laboratory of THz and Millimeter Waves, City University of Hong Kong, Hong Kong, 999077, China**

Jin Chen, Hao Guo, Zhi Chao Chen, Kam Man Shum, Ka Fai Chan, & Chi Hou Chan

**School of Information Science and Engineering, State Key Laboratory of Millimeter Waves, Southeast University, Nanjing, 210096, China**

Liang Gao, Kang Jie Lin

**Department of Electrical Engineering, City University of Hong Kong, Hong Kong, 999077, China**

Chi Hou Chan

Contributions

C.H.C., L. G, and H. G initiated the plan and supervised the entire study. C.H.C. and J.C. conceived the idea of this work. J. C. carried out the lens design and implemented the theoretical analysis. L. G designed the THz radiator array. H. G and K. J. L.

completed the testing of the THz radiator array. J.C., H.G., K.S.M., and K.F.C. designed the experimental setup. J.C. and H.G. performed the imaging experiments. Z. C. C. performed the post-processing of the original imaging data. J.C. analysed the data and prepared the manuscript with input from all authors. C.H.C. L. G, and H. G reviewed and edited the manuscript.

Corresponding author

Correspondence to Hao Guo, Liang Gao, and Chi Hou Chan.

**Ethics declarations**

Competing interests

The authors declare no competing interests.

An integrated super-resolution THz 3D imaging system based on a linear nonlocal achromatic freeform Bessel beam lens and high-power oscillator–radiator array

## Supplementary Note 1: Theoretical analysis of local and nonlocal GRIN lens

The electric field distributions of the local GRIN Bessel beam lens shown in Fig. 2a are calculated using scalar diffraction theory based on the angular spectrum method, as we have done in reference 4 and listed in Fig. S1.

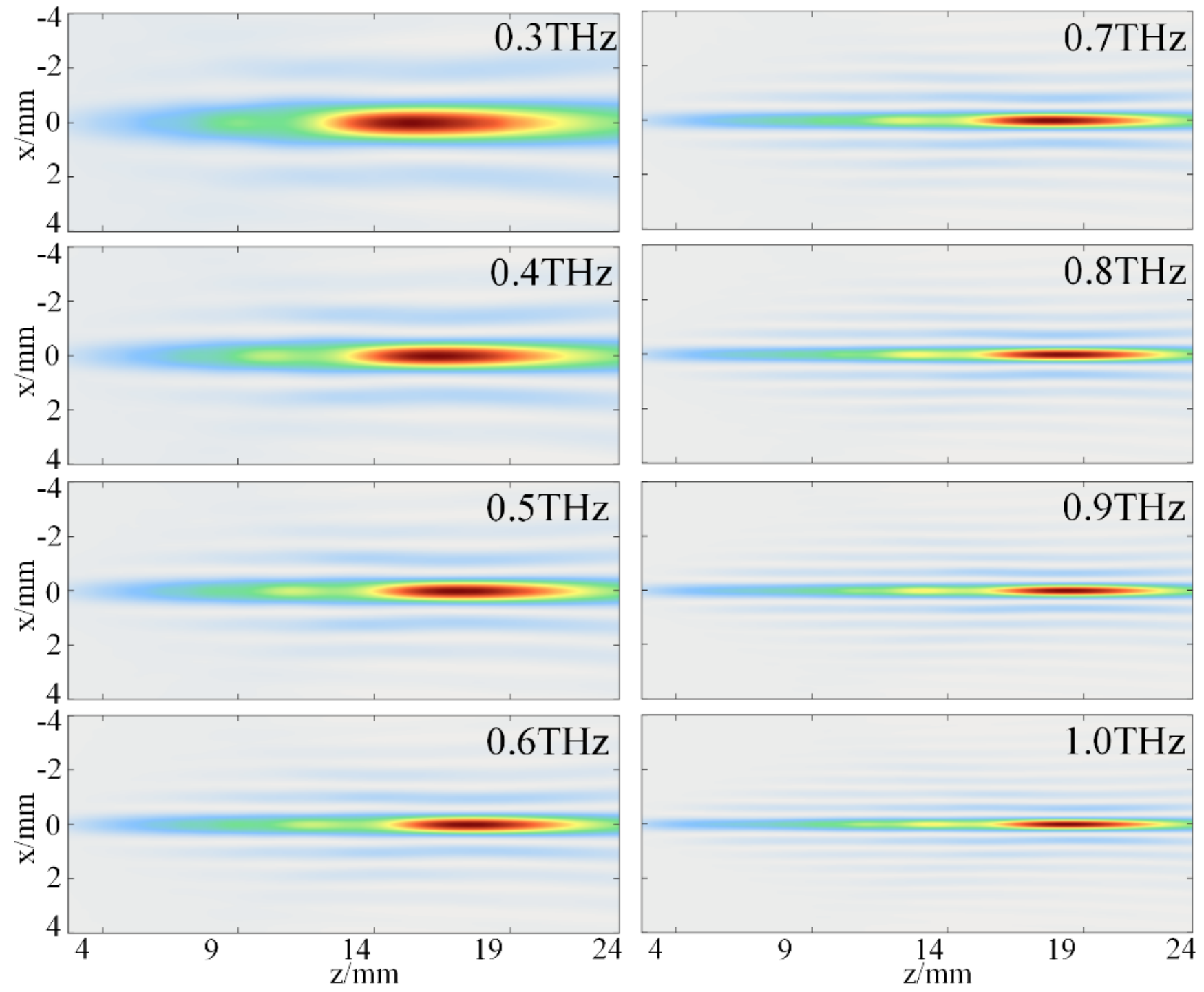

**Fig. S1 Calculated electric field distributions of the local GRIN Bessel beam lens from 0.3 to 1.0 THz.**

Fig. S2 illustrates the electric field distributions of the proposed nonlocal GRIN Bessel beam lens, the architecture of which is depicted in Fig. 2a.

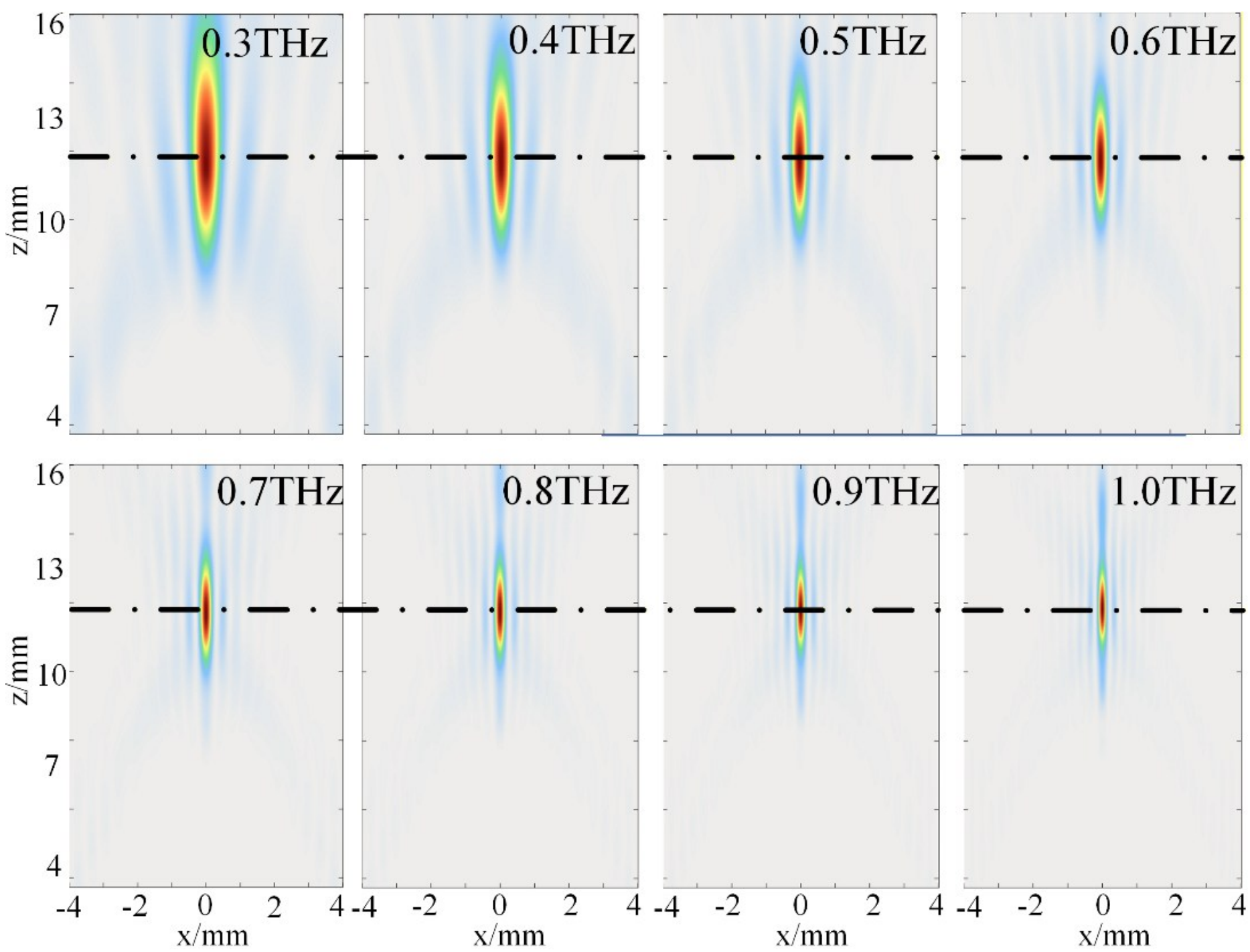

**Fig. S2 Calculated electric field distributions of our proposed nonlocal GRIN Bessel beam lens.**

To quantitatively make a comparison between the nonlocal GRIN lens and the local GRIN lens for Bessel beam generation, the extracted FWHM of each is presented in Fig. S3.

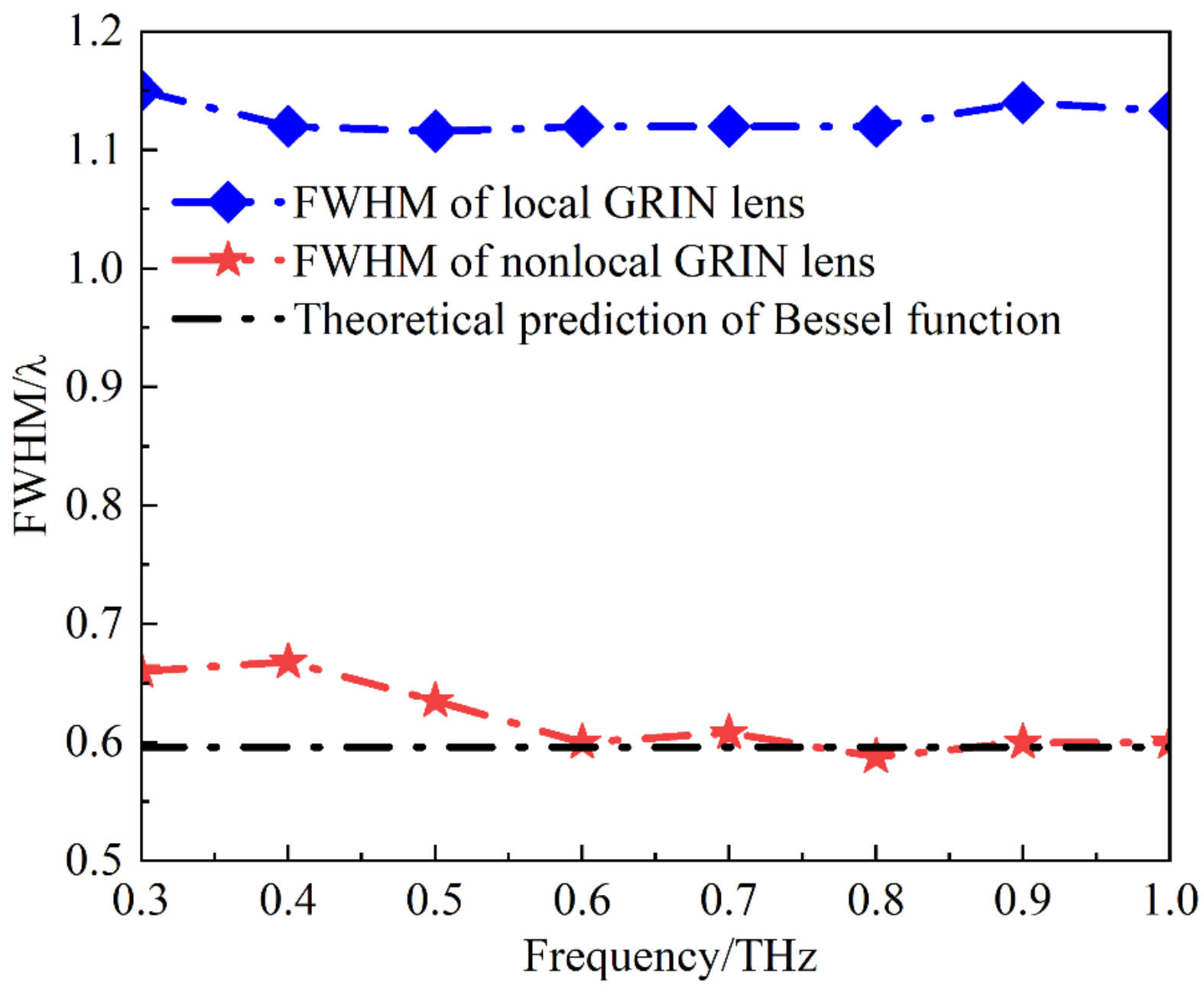

**Fig. S3 Comparison of FWHM between the local one and the nonlocal one**

The propagation equation for electromagnetic waves within the GRIN medium is the Helmholtz equation as follows:

$$\nabla^2 E + k^2 E = 0 \tag{5}$$

Since the Bessel beam is cylindrically symmetric, the Helmholtz equation under cylindrical coordinates is rewritten as follows:

$$\left(\frac{\partial}{\partial \rho^2} + \frac{1}{\rho}\frac{\partial}{\partial \rho} + \frac{1}{\rho^2}\frac{\partial^2}{\partial \phi^2} + \frac{\partial}{\partial z^2} + k^2\right) E(\rho, \phi, z) = 0 \tag{6}$$

Assuming that the solution of the above equation is composed of 3 independent parts as follows:

$$E(r, \phi, z) = R(r)\Phi(\phi)Z(z) \tag{7}$$

Subsequently, Equation 6 could be expressed as:

$$\frac{1}{R}\frac{1}{r}\frac{1}{dr}\left(r\frac{dR}{dr}\right) + \frac{1}{r^2}\frac{1}{\Phi}\frac{d^2\Phi}{d\phi^2} + \frac{1}{Z}\frac{d^2 Z}{dz^2} + k^2 = 0 \tag{8}$$

Let the azimuth part be a constant:

$$\frac{1}{\Phi}\frac{d^2\Phi}{d\phi^2} = -m^2 \tag{9}$$

The solution of Equation 9 is:

$$\Phi(\phi) = e^{im\phi} \quad (m = 0, \pm 1, \pm 2, \ldots\ldots\ldots) \tag{10}$$

For an axisymmetric Bessel beam, m = 0, it can be deduced that $\Phi(\phi) = 1$. The propagation part is also defined as a constant:

$$\frac{1}{Z}\frac{d^2Z}{dz^2} = -k_z^2 \tag{11}$$

The solution of Equation 11 is:

$$Z(z) = e^{ik_z z} \tag{12}$$

For the radial part:

$$\frac{d^2R}{dr^2} + \frac{1}{r}\frac{dR}{dr} + k_r^2 R = 0 \quad (\mathrm{k}_r = \sqrt{k^2 - k_z^2}) \tag{13}$$

The solution of Equation 13 is:

$$R(r) = J_0(k_r r) \tag{14}$$

In which $J_0$ is the Bessel function of the first kind of order zero, $k_r = ksin(\theta)$, $k_z = kcos(\theta)$, and $\theta$ is the cone angle of the Bessel beam. Finally, the electric field distribution of the Bessel beam is:

$$E(r,z) = E_0 \cdot J_0(k_r r) \cdot e^{ik_z z} \tag{15}$$

To derive the FWHM of the ideal Bessel beam, the problem is transformed into solving the following equation:

$$J_0^2(k_r r) = 0.5 \tag{16}$$

Therefore, the FWHM could be derived by:

$$FWHM \approx \frac{2\times 1.126}{k}\frac{1}{sin\theta} \approx \frac{2.252}{2\pi}\frac{\lambda}{sin\theta} \approx 0.358\frac{\lambda}{sin\theta} \tag{17}$$

**Fig. S4 Calculated electric field distributions of our proposed nonlocal freeform lens in the x-z plane**

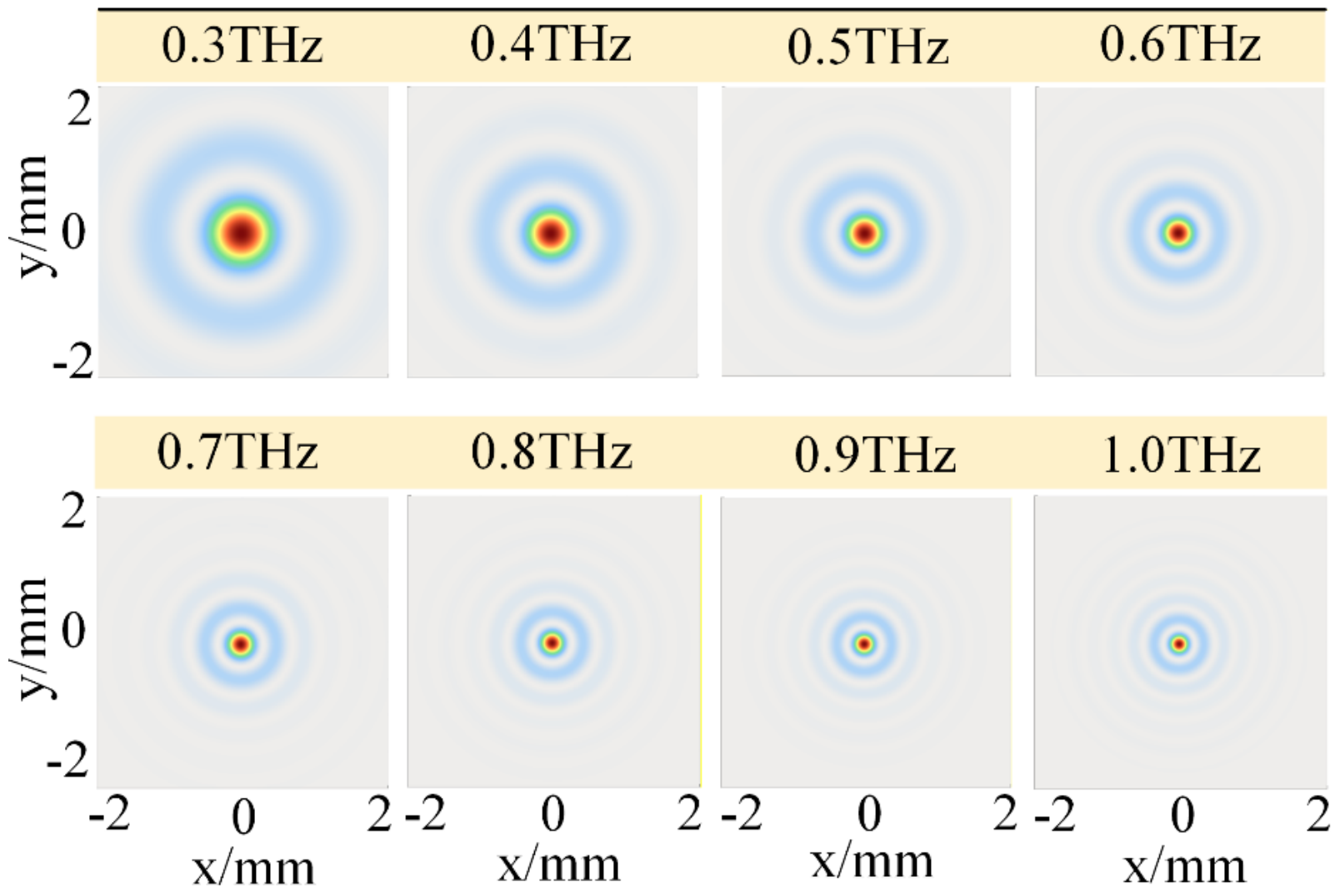

**Fig. S5 Calculated electric field distributions of our proposed nonlocal freeform lens in the x-y plane**

## Supplementary Note 2: Experimental details of the integrated THz imaging system

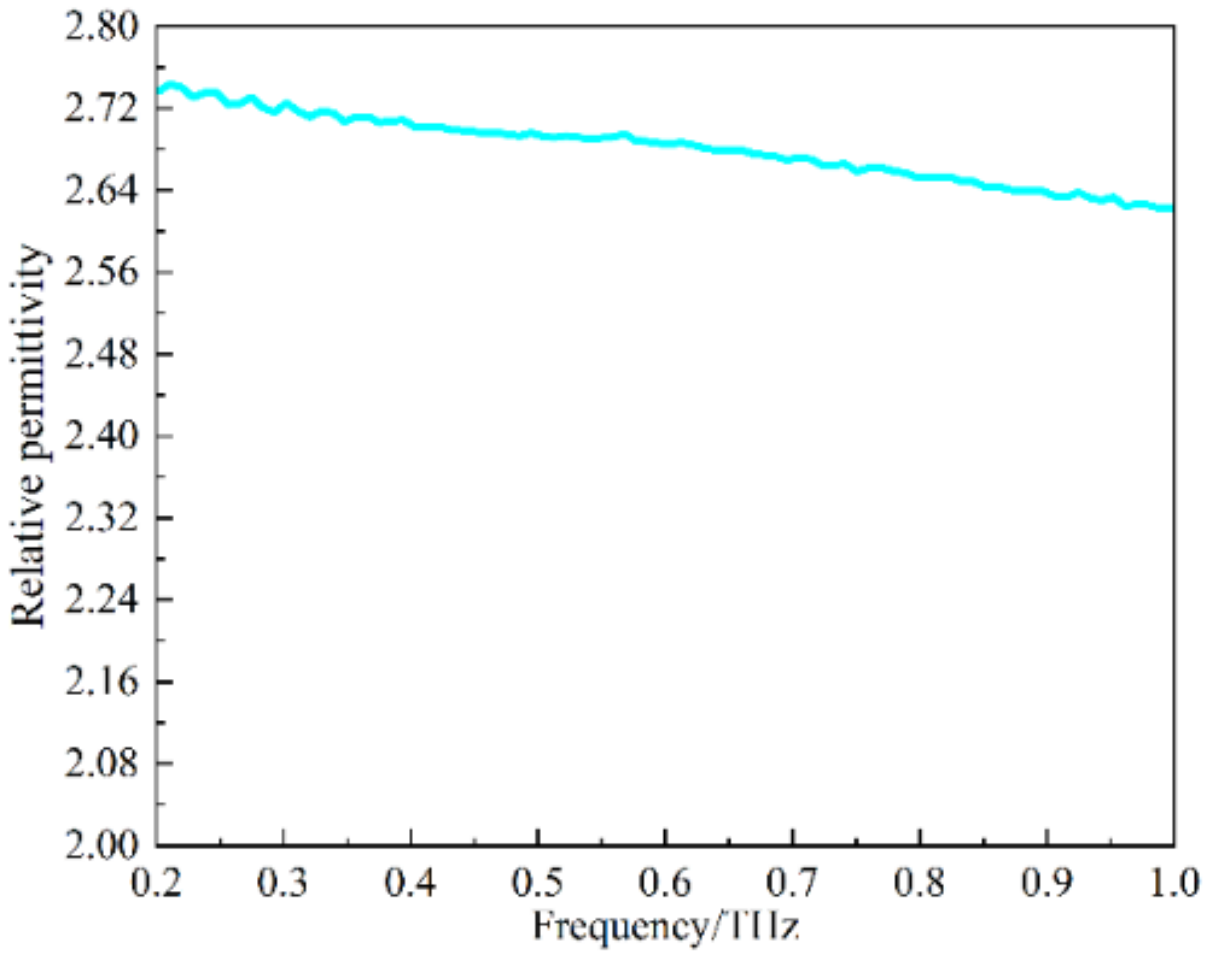

**Fig. S6.** Relative permittivity of the adopted high-temperature resin from Formlabs.

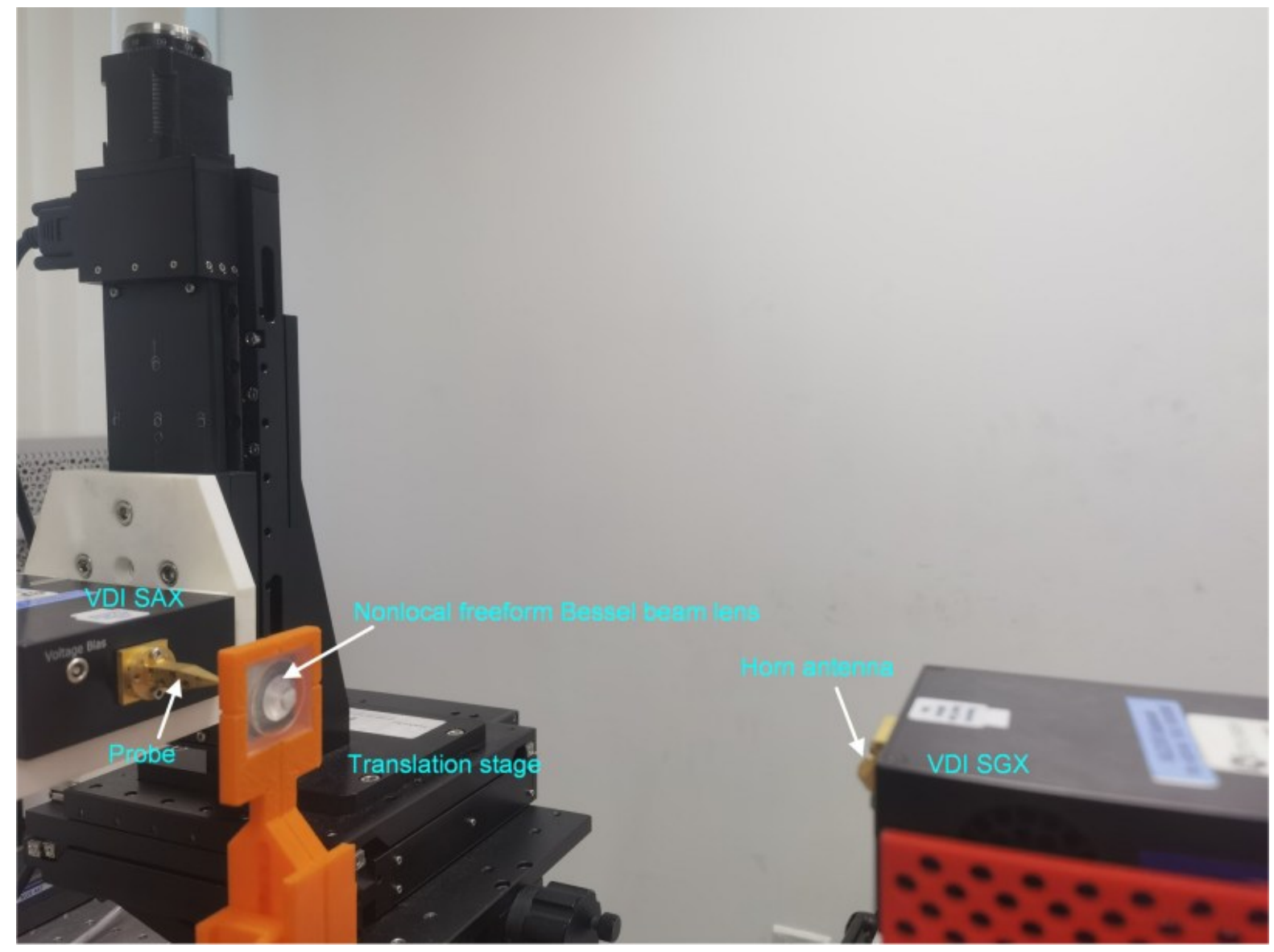

**Fig. S7 Experimental setup for measuring achromatic Bessel beam**

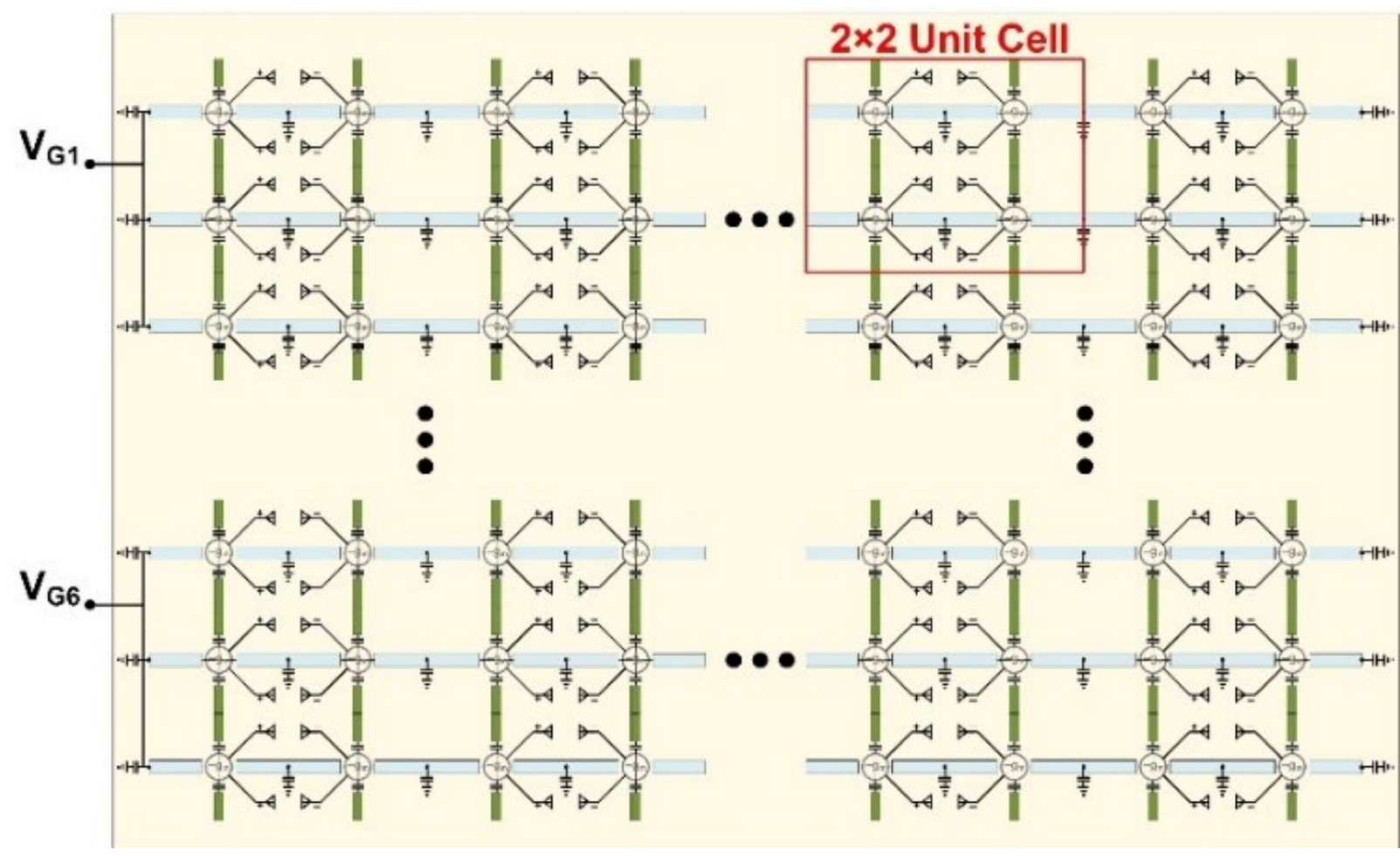

**Fig. S8 Schematic of the 18 × 16 array**

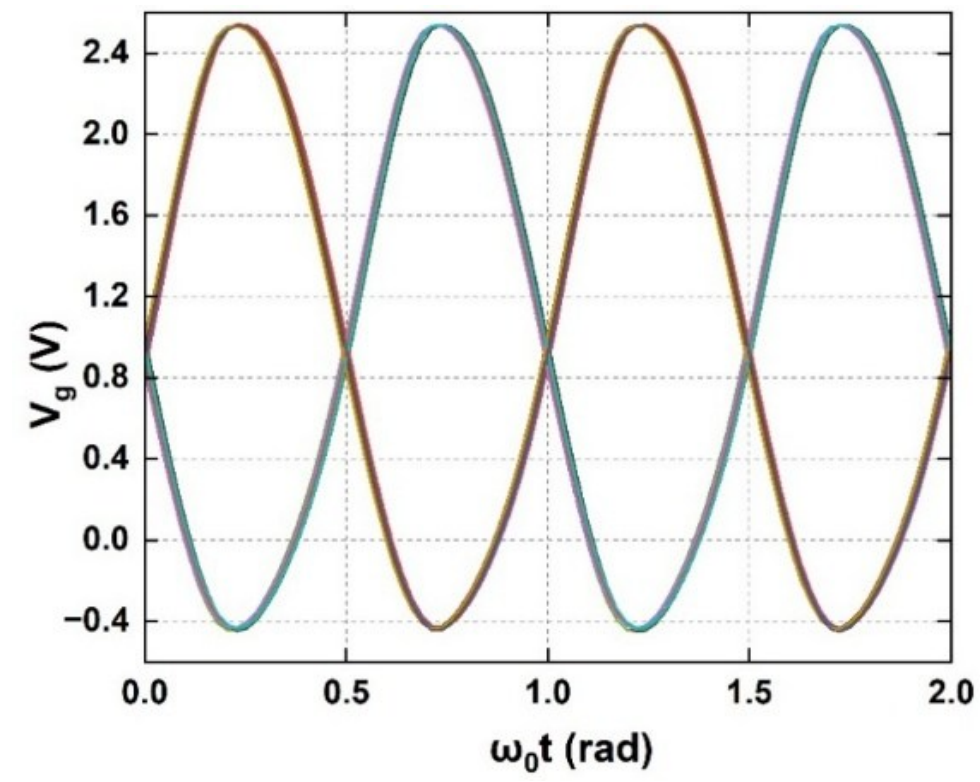

**Fig. S9 Simulated waveforms of the array.**

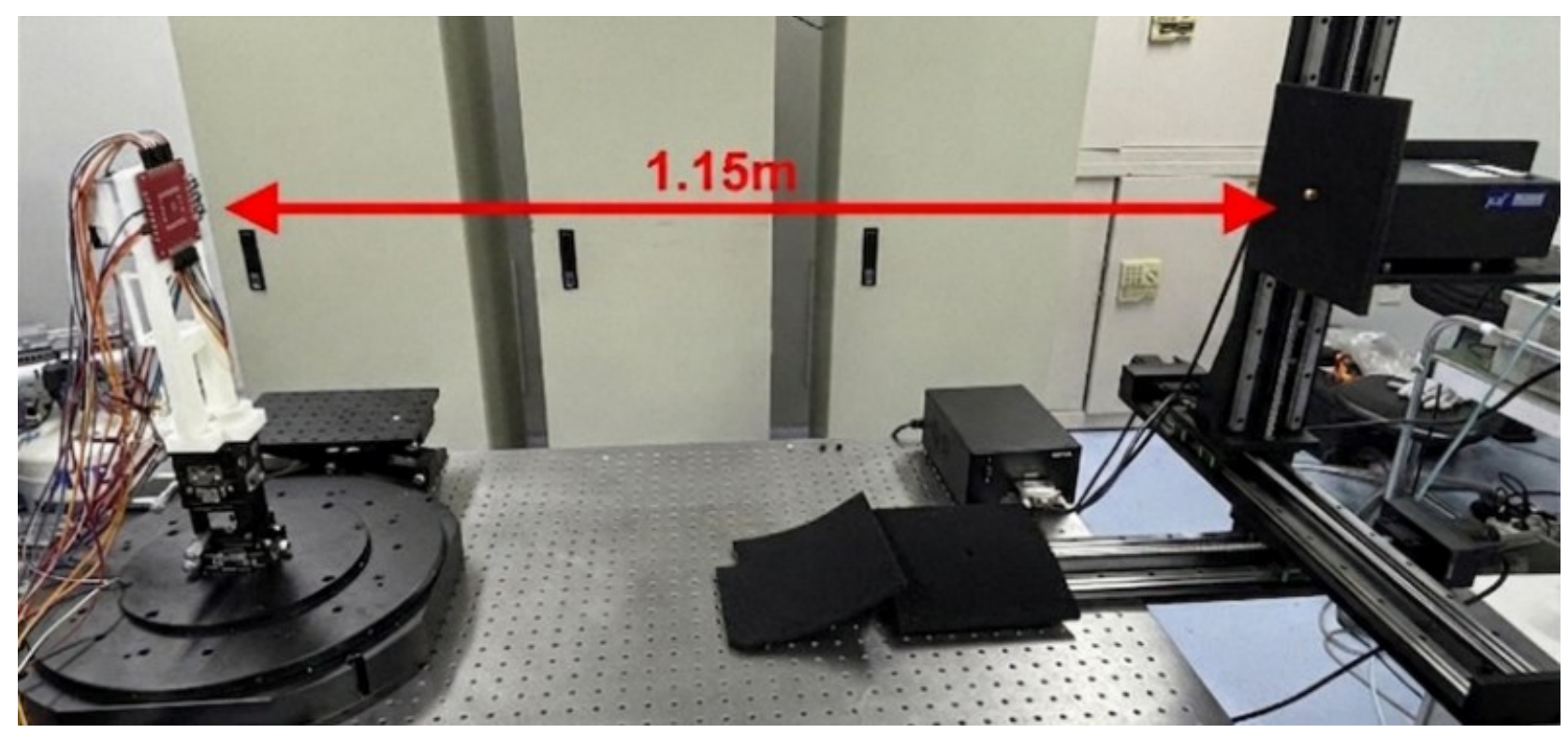

**Fig. S10 Measurement setup for THz oscillator–radiator array**

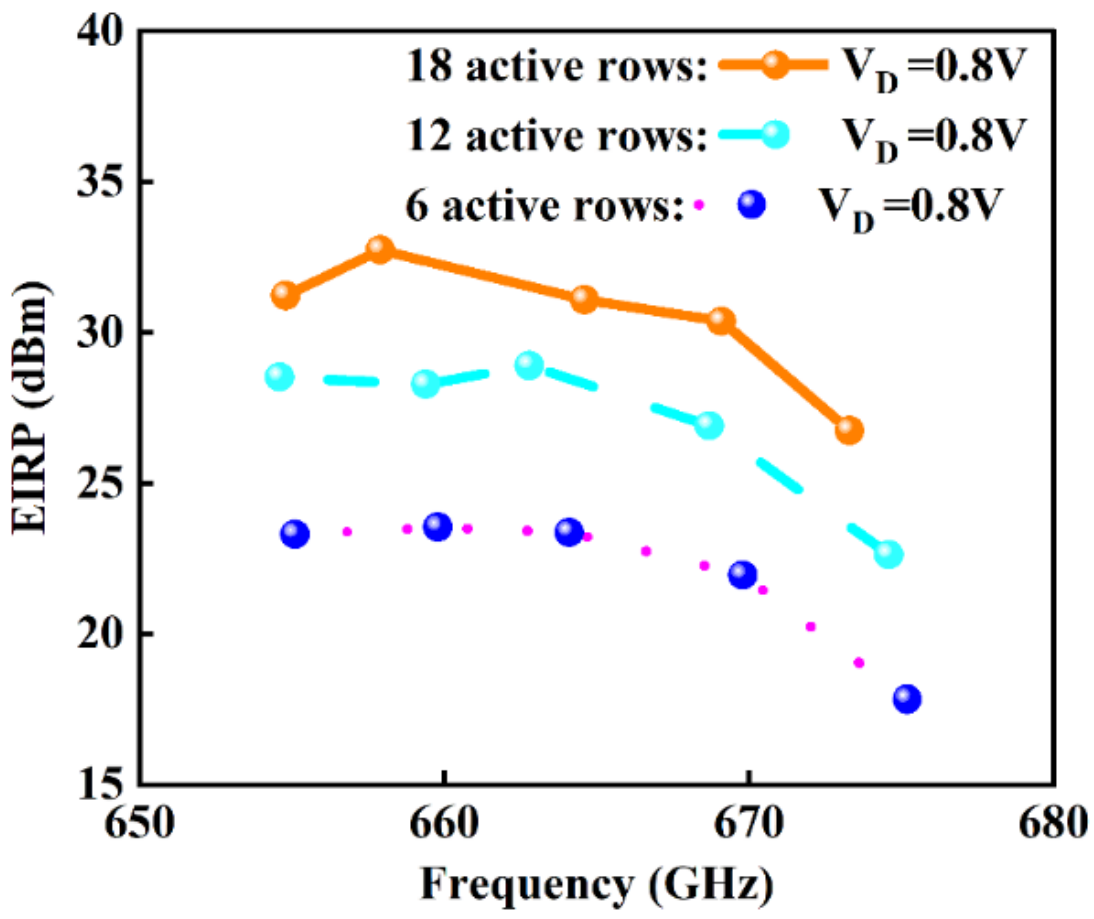

**Fig. S11 Output power of the THz oscillator–radiator array under different conditions.**

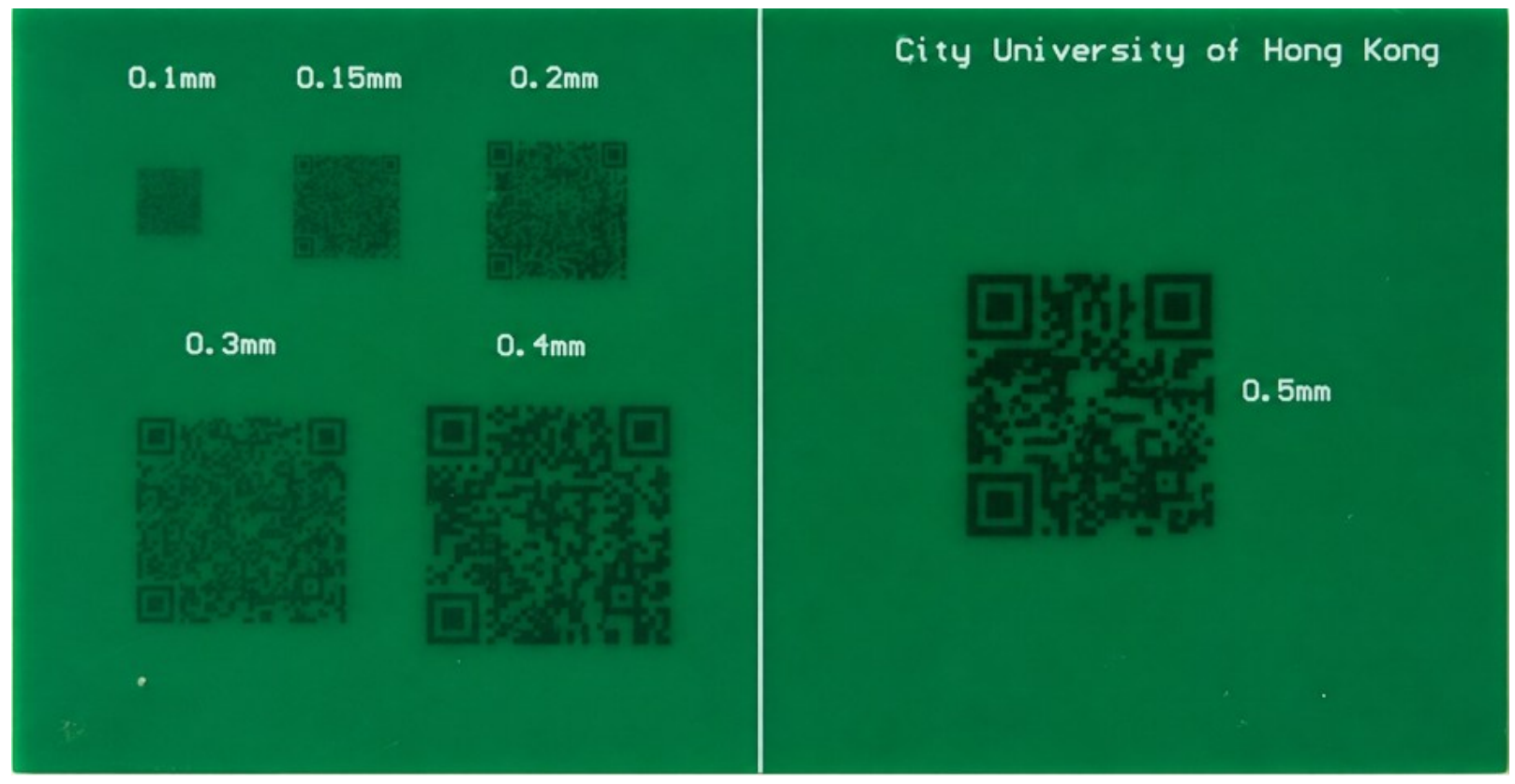

**Fig. S12 Photograph of the OR codes on the PCB panel.**

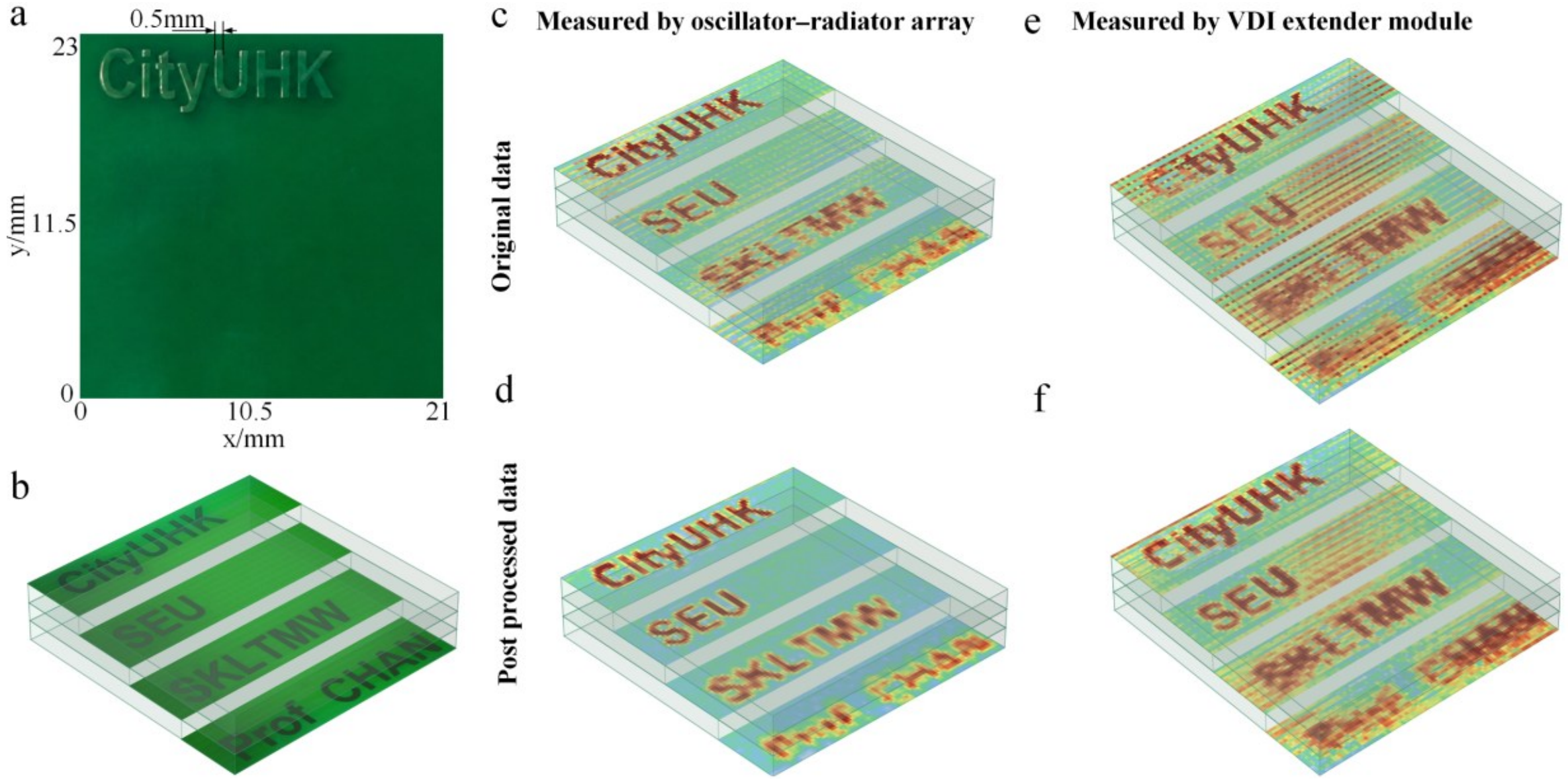

**Fig. S13 A thick Multi-layer PCB and the corresponding imaging results**. **a**. Optical photograph of this multi-layered PCB substrate. **b**. 3D geometric layout of this multi-layered PCB substrate. **c-d**. Raw data of imaging results measured by the oscillator array and the corresponding post-processed data. **e-f**. Raw data of imaging results measured by the VDI extender module and the corresponding post-processed data.

## Supplementary Note 3: Details of the post-processing technologies.

The Bessel beam spot exhibits non-negligible side lobes, so the observed image $g(x,y)$ is modeled as the linear convolution of the true transmittance $f(x,y)$ of the target with the system point spread function (PSF) $h(x,y)$, corrupted by additive noise $n(x,y)$:

$$g(x,y)=(h*f)(x,y)+n(x,y) \tag{1}$$

Where $(x,y)$ denotes the spatial coordinates on the image plane and $*$ denotes two-dimensional convolution. Sidelobes induce ring-shaped artifacts surrounding the main lobe, resulting in edge blurring and ghosting effects in the reconstructed image. To mitigate these artifacts—including sidelobe interference, periodic striping, and off-grid textures arising from deconvolution—we implement a dedicated processing pipeline comprising: frequency-domain stripe filtering, grid-aligned upsampling, TV-regularized Richardson–Lucy deconvolution, axial projection constraints, and resampling to the native grid.

Owing to the periodic internal structure of the target or the mechanical periodicity of the scanner, the raw image $g$ often contains nearly periodic stripes along the horizontal and vertical directions. In the two-dimensional Fourier spectrum, these stripes appear as discrete peaks along the frequency axes:

$$G(u,v)=F\{g\}(u,v) \tag{2}$$

Here, $G(u,v)$ denotes the Fourier spectrum of g, with $(u,v)$ representing the spatial frequency coordinates conjugate to the spatial domain coordinates $(x,y)$. The algorithm identifies the fundamental stripe frequency and its primary harmonics proximal to the $u$ - and $v$-axes, applying narrow-band stop filters at these loci to derive the notch transfer function $H(u,v)$. Defining the composite operation of detection, notch-filtering, and inverse Fourier transformation as the operator $S$, the destriped image is obtained via

$$g_f(x,y)=S\{g\}(x,y)=F^{-1}\{G(u,v)H(u,v)\} \tag{3}$$

where $g_f$ is the destriped image. Owing to the confinement of the notch filters to narrow bands proximate to the frequency axes, the core spectral content remains largely unaltered. Consequently, the underlying structural integrity of the image is preserved while the periodic stripe noise is effectively suppressed.

Before deconvolution, the physical pixel scales of the image and the point spread function (PSF) must be aligned. Let $p_{img}$ and $p_{psf}$ denote the scanning steps of the image and the PSF, respectively, with $p_{img} < p_{psf}$ in our setup. Resampling the PSF onto the coarser image grid would excessively attenuate the fine sidelobe structure. Instead, we adopt a reverse strategy: upsampling the image to match the higher-resolution PSF grid. The integer upsampling factor

$$N=round\left(\frac{p_{img}}{p_{psf}}\right) \tag{4}$$

defines the upsampling ratio, and a bicubic interpolation operator $U_N$ is applied to the destriped image at factor $N$ so that the equivalent pixel size after upsampling matches that of the PSF:

$$g_{\uparrow}(x,y)=U_N\{g_f\}(x,y) \tag{5}$$

Here $g_{\uparrow}$ is the upsampled image. Since the image and the PSF now share the same physical pixel scale, the convolution model is physically consistent on the upsampled grid and reads

$$g_{\uparrow}(x,y)=(h*f_{\uparrow})(x,y)+n_{\uparrow}(x,y) \tag{6}$$

where $f_{\uparrow}$ is the high-resolution transmittance to be recovered and $n_{\uparrow}$ is the equivalent noise on the upsampled grid.

Under a Poisson noise assumption, maximum-likelihood estimation leads to the Richardson–Lucy (RL) multiplicative iteration[i,ii]

$$\hat{f}^{(k+1)} = \hat{f}^{(k)} \cdot \left[ h * \frac{g_{\uparrow}}{h * \hat{f}^{(k)} + \varepsilon} \right] \tag{7}$$

where $\hat{f}^{(k)}$ is the estimate $f_{\uparrow}$ at the $k$-th iteration, $h(x,y) = h(-x,-y)$ is the centrally flipped PSF, and $\varepsilon > 0$ is a small constant that prevents division by zero. The RL algorithm inherently preserves non-negativity and energy conservation, exhibiting robust convergence even with complex, sidelobe-rich PSFs. However, in the absence of regularization, the iterative process progressively amplifies high-frequency noise. We therefore introduce a total variation (TV) prior[iii]

$$R_{\mathrm{TV}}(f) = \iint \|\nabla f(x,y)\| dxdy \tag{8}$$

Minimizing the negative log-posterior in the maximum a posteriori framework yields the TV-regularized RL update[iv]

$$\hat{f}^{(k+1)} = \frac{\hat{f}^{k}}{1 - \lambda \nabla \cdot \dfrac{\nabla \hat{f}^{(k)}}{\left\|\nabla \hat{f}^{k}\right\| + \eta}} \bullet \left[ h * \frac{g_{\uparrow}}{h * \hat{f}^{(k)} + \varepsilon} \right] \tag{9}$$

where $\lambda \geq 0$ controls the trade-off between smoothness and data fidelity, and $\eta > 0$ avoids the singularity when $\left\|\nabla \hat{f}^{(k)}\right\|$ approaches zero. To balance computational efficiency and stability, the divergence term in the denominator is updated only every few iterations, while the remaining steps reduce to the standard RL update. The algorithm runs for several iterations and automatically selects the optimal stopping iteration $k^*$ based on metrics such as image contrast and edge sharpness, and outputs $\hat{f}^{(k^*)}$ as the deconvolution result. The full TV-regularized RL deconvolution, parameterized by the PSF $h$, regularization weight $\lambda$, and stopping iteration $k^*$, is denoted by the operator $R_{h,\lambda,k^*}$.

Because the upsampled grid satisfies $p_{\uparrow} < p_{img}$, after many RL iterations $\hat{f}^{(k)}$ may result in oblique high-frequency textures between the original pixels that do not exist in the true scene. Such textures are essentially spurious sub-pixel patterns and tend to degrade into jagged artifacts after downsampling. To suppress them, we introduce an axial projection constraint that penalizes diagonal gradients while leaving horizontal and vertical gradients untouched. Let $D_+$ and $D_-$ denote the discrete directional derivative operators along the two diagonal directions, oriented at +45° and -45° with respect to the horizontal axis. The corresponding axial projection regularizer is defined as

$$R_{axis}(f) = \iint \left( |D_+ f|^2 + |D_- f|^2 \right) dxdy \tag{10}$$

Since this regularizer excludes the horizontal and vertical derivatives, it does not penalize axis-aligned edges of the true scene and only suppresses the oblique high frequencies introduced by deconvolution. In practice, the constraint is enforced as an explicit gradient-descent step applied once every few RL iterations:

$$\hat{f} \leftarrow \hat{f} - \beta\left( D_+^{\mathrm{T}} D_+ \hat{f} + D_-^{\mathrm{T}} D_- \hat{f} \right) \tag{11}$$

where $\beta > 0$ is the constraint step size, and $D_+^{\mathrm{T}}$ and $D_-^{\mathrm{T}}$ are the discrete adjoints of $D_+$ and $D_-$. The composite operator $D_+^{\mathrm{T}} D_+ + D_-^{\mathrm{T}} D_-$ is equivalent to the diagonal component of the discrete Laplacian; This approach selectively attenuates off-axis high-frequency components while preserving structures aligned with the coordinate axes. We formalize this correction step as the operator $A_\beta$.

Once the iterations and axial corrections are complete, the solution is downsampled from the PSF grid back to the original imaging grid using $N\times N$ area averaging:

$$\hat{f}_{out}(x,y) = \frac{1}{N^2} \sum_{(i,j)\in\Omega_{N(x,y)}} \hat{f}_{\uparrow}(i,j) \tag{12}$$

where $\hat{f}_{\uparrow}$ is the deconvolved result on the upsampled grid, $\Omega_{N(x,y)}$ is the set of indices in the $N\times N$ neighborhood on the upsampled grid corresponding to pixel $(x,y)$ on the original grid, and $\hat{f}_{out}$ is the final output image. This down-sampling step is denoted by the operator $D_{\mathrm{N}}$.

In summary, the entire pipeline can be written as the operator cascade

$$\hat{f} = D_N \circ A_\beta \circ R_{h,\lambda,k^*} \circ U_N \circ S(g) \tag{13}$$

where $S$ is frequency-domain stripe filtering, $U_N$ is integer-factor upsampling, $R_{h,\lambda,k^*}$ is TV-regularized Richardson–Lucy deconvolution, and $A_\beta$ is the axial projection constraint.

---

[i] Richardson W H. Bayesian-based iterative method of image restoration. Journal of the Optical Society of America, 62, 55-59 (1972)

[ii] Lucy L B. An iterative technique for the rectification of observed distributions. Astronomical Journal, 79, 745 (1974)

[iii] Rudin L I, Osher S, Fatemi E. Nonlinear total variation-based noise removal algorithms. Physica D: nonlinear phenomena. 60, 259-268 (1992)

[iv] Dey N, Blanc-Feraud L, Zimmer C, et al. Richardson–Lucy algorithm with total variation regularization for 3D confocal microscope deconvolution. Microscopy research and technique. 69, 260-266 (2006)